\documentclass[aps,prb,twocolumn,groupedaddress]{revtex4-1}
\usepackage{amsmath}
\usepackage{dcolumn}
\usepackage{amssymb}
\usepackage{amsfonts}
\usepackage{graphicx}
\usepackage[T1]{fontenc}
\usepackage{indentfirst}
\usepackage{color}
\usepackage{bbm}
\usepackage[normalem]{ulem}
\usepackage{xcolor}

\begin{document}
\title{ Spin-1/2 kagome Heisenberg antiferromagnet: Machine learning discovery of the spinon pair density wave ground state}
\author{Tanja \DJ uri\'c}
\affiliation{School of Physical and Mathematical Sciences, Nanyang Technological University, 21 Nanyang Link, Singapore 637371, Singapore}
\author{Jia Hui Chung}
\affiliation{School of Physical and Mathematical Sciences, Nanyang Technological University, 21 Nanyang Link, Singapore 637371, Singapore}
\author{Bo Yang}
\affiliation{School of Physical and Mathematical Sciences, Nanyang Technological University, 21 Nanyang Link, Singapore 637371, Singapore}
\author{Pinaki Sengupta}
\affiliation{School of Physical and Mathematical Sciences, Nanyang Technological University, 21 Nanyang Link, Singapore 637371, Singapore}

\date{\today}
\begin{abstract}
Spin-1/2 kagome antiferromagnet (AFM) is one of the most studied models in frustrated magnetism since it is a promising candidate to host exotic spin liquid states. However, despite numerous studies using both analytical and numerical approaches, the nature of the ground state and low-energy excitations in this system remain elusive. This is related to the difficulty in determining the spin gap in various calculations. We present the results of our investigation of the Kagome AFM using the recently developed group equivariant convolutional neural networks – a novel machine learning technique for studying strongly frustrated models. The approach, combined with variational Monte Carlo, introduces significant improvement of the achievable results accuracy for frustrated spin systems in comparison with approaches based on other neural network architectures. Contrary to the results obtained previously with various methods, that predicted $Z_2$ or U(1) Dirac spin liquid states, our results strongly indicate that the ground state of the kagome lattice antiferromagnet is a spinon pair density wave that does not break time-reversal symmetry or any of the lattice symmetries. The found state appears due to the spinon Cooper pairing instability close to two Dirac points in the spinon energy spectrum and resembles the pair density wave state studied previously in the context of underdoped cuprate superconductors in connection with the pseudogap phase. The state has significantly lower energy than the lowest energy states found by the SU(2) symmetric density matrix renormalization group calculations and other methods.

\end{abstract}

\maketitle
\section{Introduction}
\label{sec:Introduction}

Geometrically frustrated quantum magnets showed to be promising systems for discovering new phases of matter.\cite{Moessner,Lhuillier,Schmidt} Particularly interesting phases appearing in frustrated systems are various spin liquids characterized by long-range quantum entanglement, topological order and excitations with fractional quantum numbers.\cite{Balents,Savary, Broholm,Zhou} In addition to their importance in the context of high temperature superconductivity\cite{Lee2} such quntum spin liquids could allow the creation of a topological qubit and have therefore been proposed as promising platforms for topological quantum computation.\cite{Semeghini} Kagome lattice antiferromagnets are likely candidates to host various exotic spin liquid states. However, despite numerous studies kagome antiferromagnet properties still remain unclear. 

Among many numerical approaches applied to study kagome antiferromagnet properties are exact diagonalization (ED),\cite{Lauchli,Wiatek,Changlani2} density matrix renormalization group (DMRG),\cite{Jiang,Yan,Depenbrock,Jiang2,Nishimoto,Bauer,Gong,He} tensor network methods\cite{Evenbly,Xie,Mei,Niu,Liao,Haghshenas, Jiang3} and variational Monte Carlo (VMC).\cite{Ran,Iqbal,Clark2,Iqbal2,Tay,Hu,Hu2,Iqbal3,Westerhout,Fu,Kochkov,Yang} Different studies suggested so far a variety of candidate ground states. Aside from an initially proposed valence-bond crystal state with a 36-site unit cell,\cite{Evenbly,Marston,Zeng,Nikolic,Singh1,Singh2,Poilblanc,Schwandt,Poilblanc2,Hwang,Poilblanc3} majority of recent DMRG and VMC studies find spin-liquid ground states. Whilst SU(2) symmetric DMRG calculations obtain a fully gapped (nonchiral) $\mathbb{Z}_2$ topological ground state,\cite{Depenbrock} DMRG calculations supplemented with additional flux insertion indicate a gapless algebraic $U(1)$ Dirac spin liquid ground state.\cite{He} The latter is also found in the projected fermionic wave-function approaches \cite{Ran,Iqbal,Iqbal2,Iqbal3,Hermele,Ma,Iqbal4,Iqbal5} and the tensor-network states formalism.\cite{Liao,Lee}

Difficulty in finding quantum spin liquids in realistic microscopic models is largely due to lack of adequate computational methods. Whilst quantum Monte Carlo simulations face still unsolved sign problem,\cite{Pan,Troyer} tensor network and many variational approaches often favor particular kinds of states. Also, although DMRG method \cite{White,White2,Schollwock,Schollwock2} is known to be very accurate method for calculating properties of one-dimensional systems, difficulty of obtaining reliable results with DMRG greatly increases for $D$-dimensional systems with $D>1$. In addition, matrix product states within DMRG algorithm are more suitable to study gapped low-entangled states than highly entangled gapless phases.\cite{Chan}

Moreover, it has been found that two-dimensional gapless spin liquid states are particularly delicate states of matter that can be destabilized on constrained geometries like DMRG cylinders.\cite{Ferrari2} The ground-state properties for DMRG cylinders can therefore be very different from the ground-state properties for the truly two-dimensional lattices. An important step in the direction of avoiding possible destabilization of gapless spin liquid states in DMRG calculations was achieved by including fictitious magnetic fluxes in the original spin model. This flux threading through DMRG cylinders allowed detection of possible gapless $U(1)$ spin liquid ground states and Dirac points in the excitation spectrum in the kagome and triangular lattices.\cite{He,Hu3,Zhu}

In our calculations we employ novel neural-networks based machine learning (ML) techniques to study properties of the kagome lattice antiferromagnet since these techniques have so far shown great promise and potential for studying various many-body problems. An important advantage of the ML approach is suitability for implementation on graphics processing units (GPUs) that have enhanced computational capability. Namely, GPUs can perform multiple simultaneous computations and provide significant computational process speedup. Additional benefit is also powerful ML optimization libraries like NetKet\cite{Carleo2,Vicentini2} and TensorFlow\cite{Abadi} libraries.

Neural network quantum states (NQS) have recently been recognized as powerful variational wave-function ans\"atze that allow accurate representation of various quantum many-body states. Carleo and Troyer proposed an NQS ansatz based on the restricted Boltzmann machine neural network\cite{Carleo} and demonstrated it for quantum spin systems without geometrical frustration. Subsequently, NQS ans\"atze have later been extended to many complex many-body problems for which the exact solutions are still not known, such as frustrated spin systems,\cite{Cai,Liang,Choo3,Ferrari,Sharir,Hibat-Allah,Westerhout,Szabo,Nomura3,Astrakhantsev,Nomura2,Bukov,Fu,Kochkov,Yang,Roth,Roth2,Beck,Viteritti} fermionic models\cite{Nomura,Cai,Luo,Pfau,Choo,Stokes,Cassella} and topological phases of matter.\cite{Changlani,Clark,Deng,Glasser,Kaubruegger,Lu,Huang,Vieijra,Vieijra2} In addition to their initial application to determine ground state properties, NQS based  methods have been further developed to study excited energy states,\cite{Nomura3,Nomura2,Nomura,Roth,Roth2,Vieijra,Vieijra2,Duric,Choo2} time dynamics of many-body quantum systems,\cite{Schmitt,Gutierrez,Hofmann} open quantum systems\cite{Nagy,Hartmann,Vicentini,Yoshioka,Luo4,Reh2} and thermal quantum states at finite temperatures.\cite{Irikura,Nomura4}

Several recent studies showed that accuracy and performance of the NQS based methods can be significantly improved by restoring symmetries in variational NQS ans\"atze via quantum-number projections,\cite{Nomura,Mizusaki,Choo2,Misawa,Seki,Nomura2,Sharir,Hibat-Allah,Nomura3, Reh} introduction of equivariant models for various symmetry groups \cite{Pfau,Cohen,Luo2,Luo3,Bronstein,Roth,Roth2} or transformation to basis of irreducible representations for quantum states with nonabelian or anyonic symmetries.\cite{Vieijra,Vieijra2}
Equivariance based symmetrization combined with deep neural network ans\"atze demonstrated so far significant advantage in comparison to other symmetry averaging procedures allowing high-accuracy variational Monte Carlo calculations for highly frustrated quantum magnets. Notably, Roth, Szab\'o and MacDonald recently obtained very accurate results for $J_1$-$J_2$ Heisenberg models on the square and triangular lattices.\cite{Roth,Roth2}

We study the spin-1/2 kagome lattice antiferromagnet by representing the ground and low-lying excited energy states by group equivariant convolutional neural network (GCNN) ans\"atze equivariant over the full space group composed of transformations that leave the kagome lattice unchanged (lattice translations and dihedral D(6) point group transformations). GCNNs are very expressive ans\"atze that can efficiently learn from a limited number of samples. This yields to significant improvement of the achievable results accuracy within the GCNN approach combined with VMC. The GCNN and VMC approach is therefore a powerful new numerical approach for clarifying the kagome antiferromagnet properties that can provide a reference to check validity of the results obtained with other methods. 

Our numerical calculations were performed using computational speedup on GPUs and NetKet,\cite{Carleo2,Vicentini2} JAX,\cite{Frostig} FLAX,\cite{Heek} and OPTAX\cite{Hessel} libraries. Contrary to the results obtained previously with other methods, we find that the kagome lattice antiferromagnet ground state is a spinon pair density wave (PDW) state that does not break time-reversal symmetry or any of the lattice symmetries. The state appears due to the spinon Cooper pairing instability close to two Dirac nodes of the underlying $U(1)$ Dirac spin liquid found with recent flux insertion supplemented DMRG approach\cite{He} and also with projected fermionic wave-function approaches\cite{Ran,Iqbal,Iqbal2,Iqbal3,Hermele,Ma,Iqbal4,Iqbal5} and within the tensor-network states formalism.\cite{Liao,Lee} The found spinon PDW resembles the pair density wave state studied previously in the context of underdoped cuprate superconductors in connection with pseudogap phase\cite{Agterberg,Lee3,Lee4,Xu,Chakraborty} and has significantly lower energy than the ground state energy values obtained by other methods. In particular, for the studied system sizes we find that the PDW energy has $\approx 1.78\%$ lower value than the ground state energy found by SU(2) symmetric density matrix renormalization group calculations.\cite{Depenbrock}

The paper is organized as follows. In Sec. \ref{sec:GCNN} we describe the GCNN wave-function ans\"atze and explain wave-function optimization procedure. Our results for the ground state properties and low-energy excited states are presented in Sec. \ref{sec:GroundState}. Description of the found spinon pair density wave ground state  and possible induced secondary orders within Abrikosov fermion parton construction is presented in Sec. \ref{sec:PartonConstruction}. We summarize our results, draw conclusions and discuss future research directions in the last section, Sec. \ref{sec:Conclusions}.  

\section{GCNN wave-function representation and training scheme for the wave-function optimization}
\label{sec:GCNN}

Deep neural networks, and in particular deep convolutional neural networks (CNNs), have proven to be very powerful models for pattern recognition in image, video and audio data analysis,\cite{LeCun,Goodfellow,Kelleher} with both depth and convolutional weight sharing playing important role in the predictive power of these models. Convolutional weight sharing, that is, using the same weights to model for example different parts of an image, provides an effective description due to the presence of translational symmetry (up to edge effects) in most of the mentioned pattern recognition tasks. CNNs therefore use significantly smaller number of parameters than fully connected networks while preserving expressivity and capacity to learn useful features. Since quantum many-body wavefunctions for various lattice Hamiltonians similarly often possess translational symmetry, CNN based NQS ans\"atze allow efficient representation of these wavefunctions.  

CNNs can be generalized to take into account all relevant lattice symmetries. GCNNs\cite{Cohen,Kondor,Roth,Roth2} are such CNN generalizations to arbitrary nonabelian groups that take into account all space-group symmetries, including both lattice translations and point-group symmetries such as rotations and reflections. The key property of CNNs and GCNNs is equivariance of the convolution layers where increasingly abstract representations of the input are developed by alternating convolution operations with non-linearities. The equivariance property alone is sufficient to specify symmetry eigenvalues. A network (function) $f$ is equivariant with respect to a transformations $g$ if $f(gx)=g'(f(x))$ where in general $g\neq g'$ and $x$ and $g$ denote the network input and a symmetry group element, respectively. 

\begin{figure}[t!]
\includegraphics[width=\columnwidth]{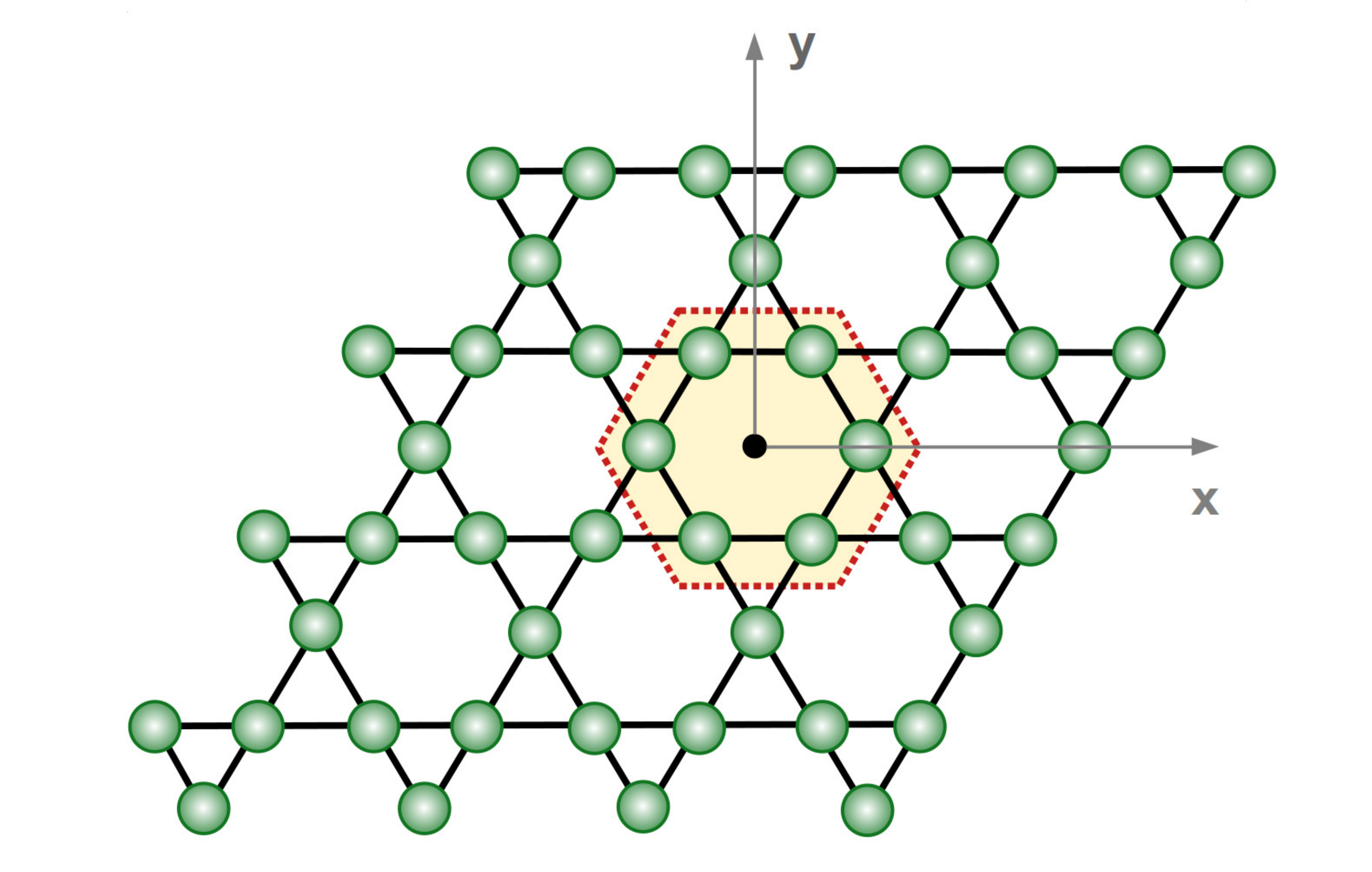}
\caption{\label{Fig:K_Kernel} Local cluster for which elements of the embedding kernels $K$ are nonzero. The $\vec{r}=(0,0)$ point is the origin for lattice sites and $D(6)$ point group symmetries.
}
\end{figure}
For systems of interacting spins on a lattice, neural network ans\"atze for the wavefunctions $|\psi\rangle$ of the lattice Hamiltonian associate a complex number $\psi(\vec{\sigma};\vec{\alpha})$ with each computational basis configuration of spins $|\vec{\sigma}\rangle$, where $\vec{\alpha}$ denotes the network parameters:
\begin{equation}\label{eq:ansatz1}
|\psi\rangle=\sum_{\vec{\left\{\sigma\right\}}}\psi\left(\vec{\sigma};\vec{\alpha}\right)|\vec{\sigma}\rangle,
\end{equation}
and $\vec{\sigma}=\left\{\sigma(\vec{r}_1),...,\sigma(\vec{r}_N)\right\}$ with $\vec{r}_1,...,\vec{r}_N$ being positions of the lattice sites. In other words, spin configurations $|\vec{\sigma}\rangle$ are the network inputs, and the complex coefficients $\psi(\vec{\sigma};\vec{\alpha})$ are the network outputs.

General GCNN architecture is described in detail in Appendix \ref{Appendix_GCNN}. For a GCNN ansatz with $N_l$ layers and with $N_f$ complex-valued feature maps (composed of $2\cdot N_f$ real-valued feature maps) in the $N_l$-th layer the network outputs $\psi(\vec{\sigma};\vec{\alpha})$ are 
\begin{equation}\label{eq:ansatz_GCNN}
\psi(\vec{\sigma};\vec{\alpha})=\sum_{\hat{g}\in G}\chi_g^*\sum_{n=1}^{N_f}\exp\left(f^{N_l}_{n,g}(\vec{\alpha})\right)
\end{equation}
where 
\begin{equation}\label{eq:GCNN_K}
\mathbf{f}_g^1=\Gamma\left(\sum_{\vec{r}}\mathbf{K}\left(\;\hat{g}^{-1}\vec{r}\;\right)\sigma\left(\;\vec{r}\;\right)\right)
\end{equation}
for a set of learnable kernels or filters $\mathbf{K}$ in the first embedding layer that generates feature maps from the input data, 
 \begin{equation}\label{eq:GCNN_W}
\mathbf{f}_g^{i+1}=\Gamma\left(\sum_{\hat{h}\in G}\mathbf{W}^i\left(\hat{h}^{-1}\hat{g}\right)\mathbf{f}_h^i\right)
\end{equation}
for a set of convolutional kernels $\mathbf{W}$ in the group convolutional layers denoted by $i=2,..,N_l$ and $\Gamma$ is scaled exponential linear unit (SELU) activation function applied separately to the real and imaginary components.

We also note that using local embedding kernels $K$, obtained by restricting nonzero $K$ elements only to local clusters, stabilizes the optimization procedure and allows numerical calculations with deeper neural networks.\cite{Roth,Roth2} The optimization procedure, i.e. the network training where the network learns from the feedback from variational principle, has faster convergence for GCNN architectures with local kernels. This is because the kernel locality structures the representation of long-range correlations and simplifies the learning landscape. Locality in the convolutional kernels $W$ is imposed by restricting the kernels to the symmetry operations of the form $\hat{g}=\hat{t}\hat{p}$ where $\hat{t}$ is a lattice translation and $\hat{p}$ is a lattice point-group symmetry operation. 

\begin{figure}[b!]
\includegraphics[width=\columnwidth]{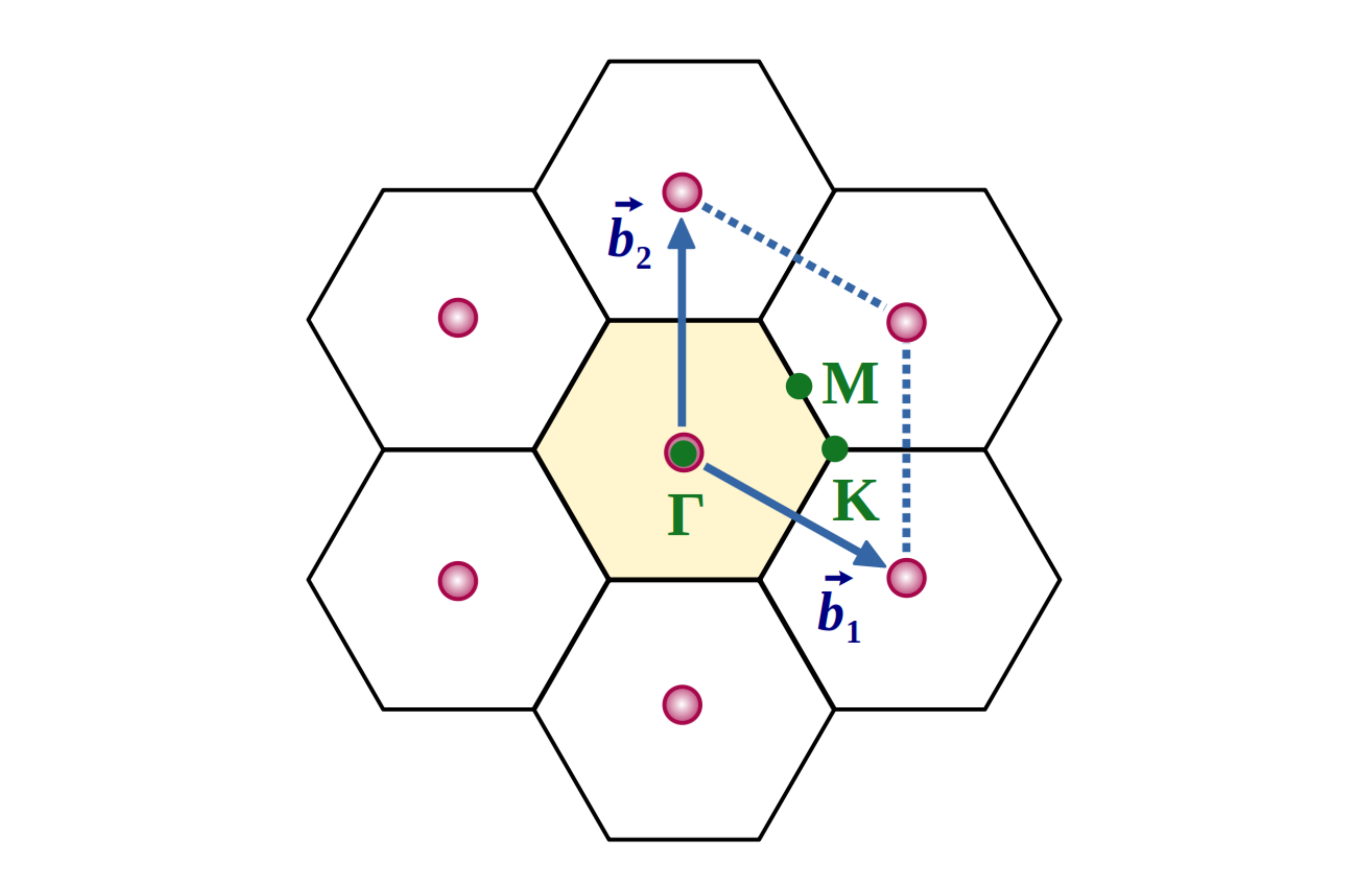}
\caption{\label{Fig:k_lattice} High symmetry points ($\Gamma$, $M$ and $K$) in the kagome lattice Brillouin zone where $\vec{b}_1$ and $\vec{b}_2$ are two primitive basis vectors in the reciprocal lattice. 
}
\end{figure}
In our calculations for the kagome antiferromagnet kernels $K$ are restricted to the symmetric local cluster illustrated in FIG. \ref{Fig:K_Kernel}. We find that choosing larger clusters with different symmetric shapes does not improve convergence nor stability of the algorithm. The set of the crystal momenta related to the lattice translations and the lattice Brillouin zone are illustrated in FIG. \ref{Fig:k_lattice}. Positions of the kagome lattice sites are $\vec{R}_{L_1,L_2,k}=L_1\cdot\vec{a}_1+L_2\cdot\vec{a}_2+\vec{r}_k$ where $\vec{a}_1=\left(1,0\right)$ and $\vec{a}_2=\left(1/2,\sqrt{3}/2\right)$ are two primitive vectors of the lattice unit cell containing three inequivalent sites, and $\vec{r}_k$ describe positions of the sites within the unit cell: $\vec{r}_1=\left(1/2,0\right)$, $\vec{r}_2=\left(1/4,\sqrt{3}/4\right)$ and $\vec{r}_3=\left(3/4,\sqrt{3}/4\right)$. Two primitive basis vectors $\vec{b}_1=2\pi\left(1,-1/\sqrt{3}\right)$ and $\vec{b}_2=4\pi\left(0,1/\sqrt{3}\right)$ in the reciprocal lattice are calculated as $\vec{b}_1=2\pi\frac{\hat{O}_R\vec{a}_2}{\vec{a}_1\cdot\hat{O}_R\vec{a}_2}$ and $\vec{b}_2=2\pi\frac{\hat{O}_R\vec{a}_1}{\vec{a}_2\cdot\hat{O}_R\vec{a}_1}$ where $\hat{O}_R$ represents $\pi/2$ (anti-clockwise) rotation matrix.

Our calculations have been performed with GCNN ans\"atze having 6 layers and 6 feature maps in each layer for the kagome lattice cluster with 48 lattice sites and GCNN ans\"atze having 4 layers and 4 feature maps in each layer for the kagome lattice cluster with 108 lattice sites. In addition to the space-group symmetry, composed of lattice translations and dihedral $D(6)$ point group transformations, the total symmetry group for the ans\"atze also includes $\mathbb{Z}_2$ spin parity group generated by $P_{\mathbb{Z}_2}=\prod_i\sigma_i^x$ for the eigenstates with $M_z=\sum_{i=1}^N\sigma(\vec{r}_i)=0$ where $\sigma(\vec{r}_i)\in\left\{-1/2,1/2\right\}$ and $N$ is the number of lattice sites. Irreducible representations of the spin parity group are specified by the eigenvalues of $P_{\mathbb{Z}_2}$ which can be $+1$ or $-1$. Total spin quantum number for an eigenvalue $|\psi_{P_{\mathbb{Z}_2}}\rangle$ that is also an eigenvalue of the total spin operator can be obtained by calculating $\langle \psi_{P_{\mathbb{Z}_2}}|\hat{\vec{S}}_{tot}^2ˇ|\psi_{P_{\mathbb{Z}_2}}\rangle$. The total spin operator is $\hat{\vec{S}}_{tot}=\left(\hat{S}_{tot}^x,\hat{S}_{tot}^y,\hat{S}_{tot}^z\right)$ where $\hat{S}_{tot}^q=\sum_{i=1}^N \hat{S}_i^q$ with $q\in\left\{x,y,z\right\}$ and $\hat{\vec{S}}^2_{tot}|\psi_S\rangle=S(S+1)|\psi_S\rangle$ where $|\psi_S\rangle$ is an eigenstate with total spin quantum number $S$. 

 Within the VMC approach variational wavefunctions described by GCNN ans\"atze are optimized using stochastic reconfiguration (SR) method introduced by Sorella \emph {et al.}.\cite{Sorella1,Sorella2,Sorella3,Becca} Details of the SR method are explained in the Appendix \ref{Appendix_SR}. 
The SR iterative update rule for the network parameters is 
\begin{equation}\label{eq:SR_update}
\vec{\alpha}_{t+1}=\vec{\alpha}_t-\Delta\tau G^{-1} \vec{\nabla}_{\vec{\alpha}}\mathcal{L},
\end{equation}
where $t$ denotes iteration, $\mathcal{L}$ is a loss function, $\Delta \tau$ a small imaginary time step that corresponds to the learning rate and $G$ quantum geometric tensor (QGT) metric matrix 
with $\partial_{\alpha_i}=\partial/\partial \alpha_i$ being partial derivatives with respect to the network parameters $\vec{\alpha}=\left\{\alpha_1,...,\alpha_{N_p}\right\}$.

The most common choice of the loss function in VMC is the variational energy:
\begin{equation}\label{eq:Energy_VMC}
\mathcal{L}_E\equiv E= \frac{\langle\psi\left(\vec{\alpha}\right)|H|\psi\left(\vec{\alpha}\right)\rangle}{\langle\psi\left(\vec{\alpha}\right)|\psi\left(\vec{\alpha}\right)\rangle}. 
\end{equation}
where $H$ is the Hamiltonian of the system. For the kagome lattice antiferromagnet the Heisenberg Hamiltonian is 
\begin{equation}\label{eq:H}
H=J\sum_{\langle i,j \rangle}\vec{S}_i\cdot\vec{S}_j,
\end{equation}
where the exchange coupling $J>0$ and $\langle i,j \rangle$ denotes all possible nearest-neighboring (NN) pairs of the kagome lattice sites and the spin $1/2$ operators $\hat{\vec{S}}_i=\frac{1}{2}\left(\hat{\sigma}^x,\hat{\sigma}^y,\hat{\sigma}^z\right)$
can be written in terms of the Pauli matrices $\hat{\sigma}_i$ with $i\in\left\{x,y,z\right\}$.  

Eq. (\ref{eq:SR_update}) is however ill-conditioned, meaning that a small error in the energy gradients may cause a large error in the parameter updates. Also the QGT matrix $G$ can be non-invertible. The $G$ matrix regularization is therefore required to stabilize calculations and to achieve reliable convergence. We find that adding both a scale-invariant shift $\epsilon_1$ and a constant shift $\epsilon_2$ to diagonal entries
\begin{equation}\label{eq:QGT_regularisation}
G_{ii}\rightarrow G_{ii}+\epsilon_1G_{ii}+\epsilon_2
\end{equation}
provides suitable regularization at the later stages of the calculations when the variational energy is relatively close to the ground state energy. However, the mentioned regularization scheme is often unstable at the beginning of the network training, exhibiting run-away-like instabilities typical for non-stoquastic quantum Hamiltonians with rugged variational manifold landscape.\cite{Bukov} For GCNN architectures similar behavior was previously noticed in studies of the frustrated $J_1$-$J_2$ model.\cite{Roth,Roth2}

The instability problem at the initial stages of the training can be remedied by modifying considered loss function, that is by reshaping the optimization landscape, as it was demonstrated in several recent studies considering other complex models.\cite{Roth,Roth2,Roth3,Wu,Hibat-Allah2,Hibat-Allah3} Adding a pseudo-entropy reward term with an effective temperature $T$ to the energy loss function encourages more even sampling of the Hilbert space and leads to the total loss function:
\begin{equation}\label{eq:Free_energy}
\mathcal{L}_F\equiv F=E-TS,
\end{equation}
corresponding to free energy at the temperature $T$ where the entropy is defined as\cite{Roth,Roth2,Roth3,Wu,Hibat-Allah2,Hibat-Allah3}
\begin{equation}\label{eq:entropy}
S=-\sum_{\vec{\sigma}}\frac{|\psi\left(\vec{\sigma},\vec{\alpha}\right)|^2}{\sum_{\vec{\sigma'}}|\psi\left(\vec{\sigma'},\vec{\alpha}\right)|^2}\log \frac{|\psi\left(\vec{\sigma},\vec{\alpha}\right)|^2}{\sum_{\vec{\sigma'}}|\psi\left(\vec{\sigma'},\vec{\alpha}\right)|^2},
\end{equation}
and $\psi\left(\vec{\sigma},\vec{\alpha}\right)$ is an unnormalized neural-network ansatz wavefunction. Within this approach the effective temperature is set to some initial $T=T_0$ at the start of the training and then gradually lowered to zero as the training proceeds according to a schedule function $T_n$ for the effective temperature in the $n$-th training step. As explained in Appendix \ref{Appendix_SR}, minimizing free energy corresponds to minimizing Kullback-Leibler (KL) divergence that measures the closeness between two distributions.

In our kagome lattice antiferromagnet study, the function $T_n$ of the form $T_n=T_0e^{-\lambda n}$, with $T_0=0.5$ and $\lambda=0.02$ most effectively solves the problem with initial instabilities for a range of random initial parameter values. We however note that for larger lattice sizes we still find residual instabilities for some initial random parameter values that are prone to the instabilities and lead to unstable trajectories in the network parameters manifold. In such cases the optimization procedure is restarted with different random initial parameter values that lead to stable trajectories. 

We also note that some trajectories on the parameter manifold remain in landscape saddles (local minima) that correspond to different physical state than the lowest energy state in particular symmetry sector. Different trajectories on the parameter manifold for different initial parameter values are irrelevant provided that the final optimized network parameters represent the same quantum state. Appearance of trajectories with final parameters corresponding to local minima with different energy is an indication of a rugged underlying optimization landscape. Such trajectories reflect presence of saddles in the optimization landscape that are located in deep valleys separated by high energy barriers that are very difficult for the algorithm to overcome. 

An additional signature of such highly complex optimization landscape is also drastically different performance of different optimizers, for example stochastic gradient descent, Adam,\cite{Kingma} and SR\cite{Sorella1,Sorella2,Sorella3,Becca} optimizers. In initial stages of our calculations we also tried to apply Adam algorithm to optimize energy loss function, however SR optimization yielded more accurate results. Such behaviour is further confirmation that frustrated non-stochastic character of the studied quantum Hamiltonian, that implies number of constraints on the ground state that minimizes energy, also implies a rugged energy landscape with many curved landscape saddles placed in deep valleys. The GCNN simplification of the optimization landscape is based on the fact that introduction of equivariance in the neural network ansatz, and projection on a particular irreducible representation of the relevant symmetry group that corresponds to the symmetry properties of the ansatz wavefunction, eliminates some of the local minima from the optimization landscape resulting in simplified and less rugged landscape.

The sum in Eq. (\ref{eq:Free_energy}) is over all basis states $\vec{\sigma}$ and at large values of $T$ at the beginning of the training the network tries to first learn states with similar amplitudes for all basis states. In addition to the training stabilization, this pre-training of the phases also significantly increases achievable results accuracy which is consistent with findings in several previous neural network studies.\cite{Szabo,Roth}

Within the optimization procedure we take relatively large learning rate (imaginary time step) $\Delta \tau=0.05$ in Eq.(\ref{eq:SR_update}) and $N_s=2^{12}$ or $2^{13}$ Monte Carlo samples in each training step for the clusters with 48 and 108 lattice sites respectively. The number of samples is taken to be the maximum number allowed by the available GPU memory restrictions, since training with more samples helps the GCNN ansatz to learn the wave function representations that have better generalization properties to the parts of the Hilbert space that were not sampled. Although larger learning rates can cause non-monotonic behavior of the loss function and evaluated energy values as the training proceeds, it helps the algorithm to escape from local minima and leads to higher accuracy for the final results. 

We also find noticeable sensitivity of the algorithm to the choice of the QGT regularization parameters $\epsilon_1$ and $\epsilon_2$ in Eq.(\ref{eq:QGT_regularisation}). 
In our calculations $\epsilon_1$ and $\epsilon_2$ values are taken to be $\epsilon_1=0.001$ and $\epsilon_2=0.001$. We note that taking larger values of the regularization parameter can cause decrease of the obtained results accuracy which is related to the loss function symmetry breaking due to a weak regularizer. When a loss function is invariant under a continuous symmetry it has a manifold of equivalent degenerate minima. Any of these minima is then a valid solution to the optimization problem. The minima degeneracy is however lifted by introducing a small symmetry breaking term like a weak regularizer and the model then prefers one minimum over all other minima. Symmetries can be broken either explicitly due to an additional regularization term or spontaneously by couplings between the model parameters resulting in a slow convergence after initial fast decrease of the loss function value.\cite{Bamler}

\section{Low-lying energy states and ground state properties}
\label{sec:GroundState}
\begin{figure}[t!]
\includegraphics[width=\columnwidth]{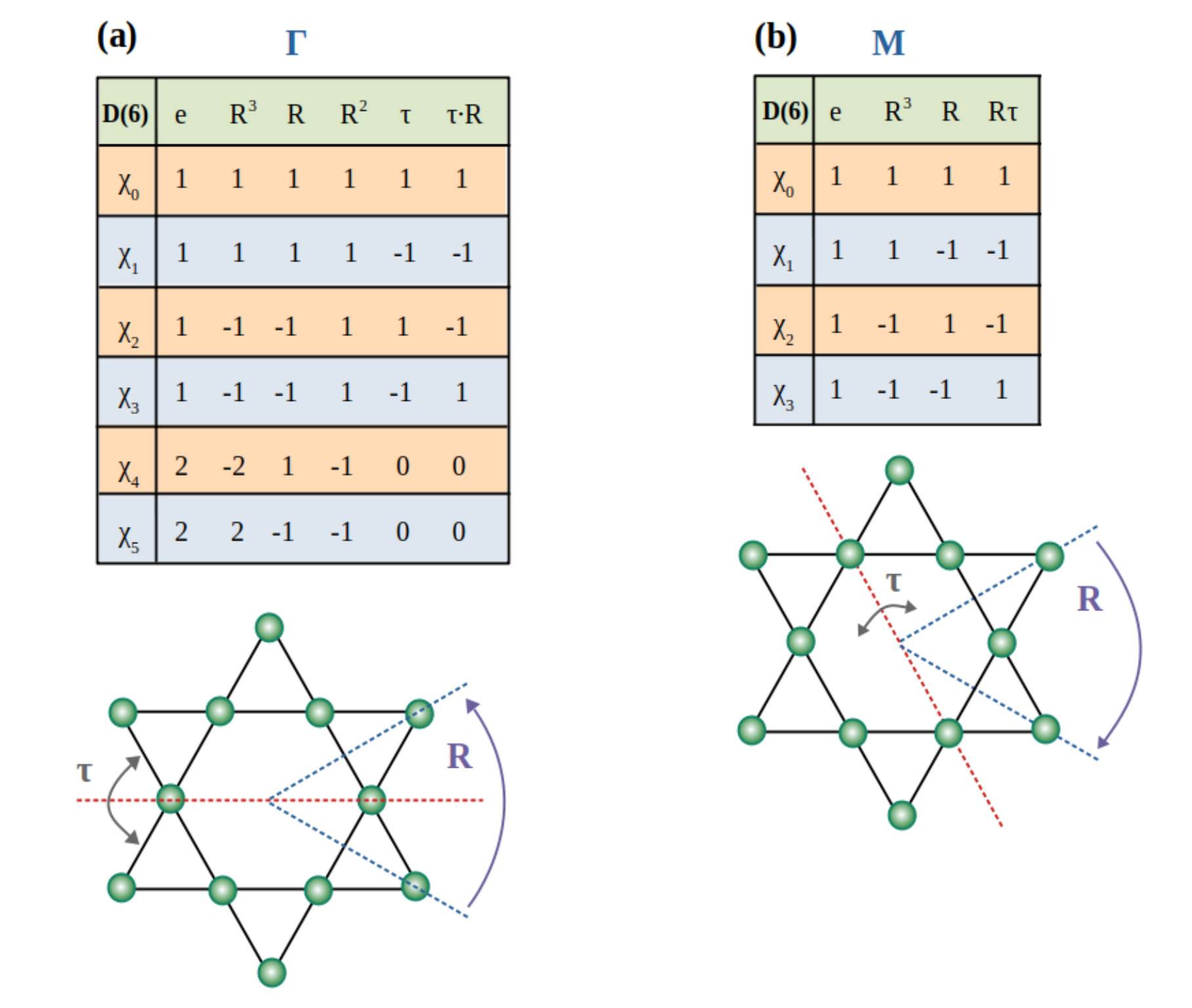}
\caption{\label{Fig:Characters_D6}Group elements of the kagome lattice little group for the crystal momentum (a) $\vec{q}=(0,0)$ ($\Gamma$ high symmetry point in the kagome lattice Brillouin zone) and (b) $\vec{q}=(\pi,\pi/\sqrt 3)$ (M high symmetry point in the kagome lattice Brillouin zone) and corresponding irreducible representation characters. Here $R$ is (anticlockwise) $\pi/3$ rotation around the center of the hexagonal plaquette and $\tau$ is reflection about the line with angle (a) zero and (b) $-\pi/3$ with respect to the $x$-axis. The little group is a subgroup of the kagome lattice point group D(6) that leaves the crystal momentum invariant.}
\end{figure}

In this section we present results for the ground state properties and low-lying eigenstates in the energy spectrum for the kagome lattice antiferromagnet clusters with 48 ($4\times 4\times 3$) and 108 ($6\times 6\times 3$) lattice sites with periodic boundary conditions. It is important to note that results for the 108 lattice sites  are of particular importance since in that case all high symmetry points are present in the lattice Brillouin zone shown in FIG. \ref{Fig:k_lattice}. The space group for $N_1\times N_2\times 3$ kagome lattice clusters with $N_1=N_2=L$ and periodic boundary conditions contains $12\times L^2$ symmetry operations obtained by combining $L^2$ lattice translations and 12 D(6) point group symmetries (rotations and reflections). The space group irreducible representations are usually constructed and described in terms of a set of symmetry related crystal momenta called star, and a subgroup of the point group called little group that leaves these symmetry related crystal momenta invariant,\cite{Aroyo} as illustrated in FIG. \ref{Fig:Characters_D6} for $\Gamma$ and $M$ high symmetry points shown in FIG. \ref{Fig:k_lattice}.

To check the accuracy of the GCNN and VMC approach we first calculate the ground state energy and singlet and triplet energy gaps for 48 sites kagome lattice cluster. The singlet excitation gap $\Delta_s=E_1(S=0)-E_0(S=0)$ is defined as energy difference of the two lowest energy states in the total spin $S=0$ sector, whilst the triplet spin excitation gap $\Delta_t=E_0(S=1)-E_0(S=0)$ can be obtained as energy difference of the lowest energy states in the total spin $S=1$ and $S=0$ sectors.

First the ground state energy is found by minimizing the free energy loss (Eq. (\ref{eq:Free_energy})) with fully symmetric wave-function ansatz and spin parity $P_{\mathbb{Z}_2}=1$ using SR optimization algorithm. Low-lying excited states are then determined using minimization of the energy loss function (\ref{eq:Energy_VMC}) with help of the transfer learning scheme, viz., the scheme to stabilize and expedite training within SR optimization for the lowest energy states in different symmetry sectors by choosing the GCNN ans\"atze for those states to have the same architecture as for the ground state. The optimized ground state parameters are used as initial guess for the parameters of the ans\"atze. The ans\"atze therefore have the same number of layers and the same number of features in each layer as the ground state ansatz, but with different space group irreducible representation characters that reflect symmetry properties of a particular low-lying energy state. The ans\"atze can in general also have different spin parity $P_{\mathbb{Z}_2}$ equal to $1$ or $-1$. Within this approach initial guess for a particular low-lying energy state wave function already has low variational energy and typical correlations characteristic for the low-energy eigenstates, resulting in more stable training yielding more accurate results for the energy eigenvalues.  

\begin{figure}[b!]
\includegraphics[width=\columnwidth]{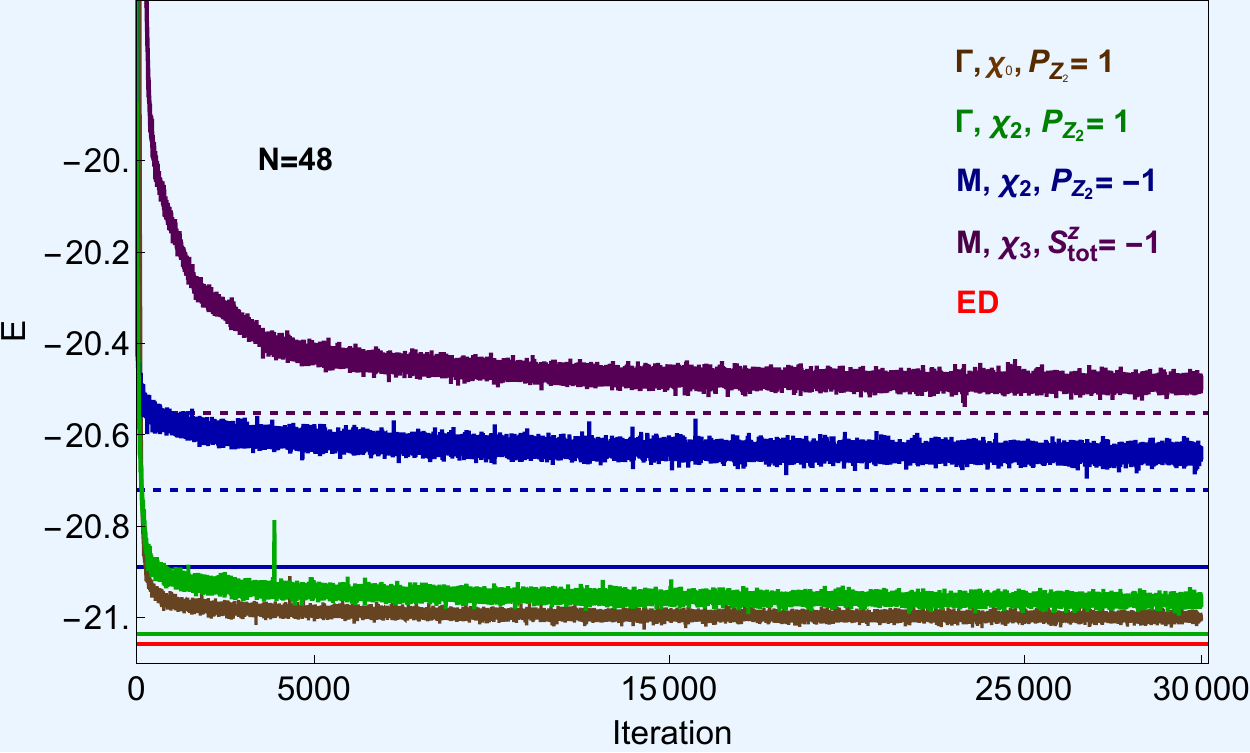}
\caption{\label{Fig:E_4x4x3}Low energy eigenstates for the kagome lattice cluster with 48 sites ($4\times4\times3$ cluster) and periodic boundary conditions. 
The ground state energy that is in the total $S^z=0$ sector (brown line graph) and the shown higher energy eigenstate in the $S^z_{tot}=-1$ sector (purple line graph) are obtained with SR minimization of the free energy loss defined in Eq. (\ref{eq:Free_energy}). The higher energy eigenstates in the $S^z_{tot}=0$ sector (green and blue line graphs) are obtained with SR minimization of the energy loss function defined in Eq. (\ref{eq:Energy_VMC}) and with help of the transfer learning scheme described in the main text. The GCNN ans\"atze for the eigenstates have 6 layers and 6 features in each layer. The space-group characters for the higher energy eigenstates in the total $S^z=0$ sector correspond to the space-group characters of the first higher energy state above the ground state in the total $S=0$ sector (green line graph) and to the lowest energy state in the total $S=1$ sector (blue line graph). The red, blue, green and purple lines denote corresponding ED results.\cite{Lauchli} For the ground and the first higher energy state in the total $S=0$ sector the SR algorithm finds global energy minima. In the symmetry sector of the lowest $S=1$ eigenstate the SR algorithm finds a local minimum corresponding to two triplons ($S=1$ excitations) forming a total $S=0$ eigenstate (dashed blue line). This finding reflects spinon Cooper pairing instability explained further in the main text. For the $S_z^{tot}=-1$ sector the algorithm finds a local minimum that corresponds to three triplons (dashed purple line) and provides a better estimate for the triplet energy gap. 
}
\end{figure}
The energies $E=\langle \psi |H|\psi\rangle/\langle\psi|\psi\rangle $ obtained for 48 lattice sites ($4\times 4\times 3$ cluster) with periodic boundary conditions at each training step with the free energy loss (Eq. (\ref{eq:Free_energy})) are shown in FIG. \ref{Fig:E_4x4x3} and converged energy values correspond to eigenvalues for the studied low energy states. The results demonstrate that GCNN approach with the ansatz composed of only 6 layers and with 6 features in each layer already allows the ground state energy accuracy with the relative error $R$ of order $10^{-3}$ with respect to the ED result ($R=(E_{\chi_0^0}^{GCNN}-E_{\chi_0^0}^{ED})/E_{\chi_0^0}^{ED} \approx 0.3 \%$ with $E_{\chi_0^0}^{ED}=-21.057787063$ and $E_{\chi_0^0}^{GCNN}=-21.00\pm 0.01$). As for the ground state, the result for the first higher energy state above the ground state in the total $S=0$ sector clearly demonstrates ability of the algorithm to escape local minima and slow convergence due to a weak breaking of the continuous SU(2) symmetry. Obtained value of the singlet energy gap $\Delta_s^{GCNN}= 0.03\pm 0.01$ is in good agreement with the ED result $\Delta_s^{ED}=0.021217$. 

\begin{figure}[t!]
\includegraphics[width=\columnwidth]{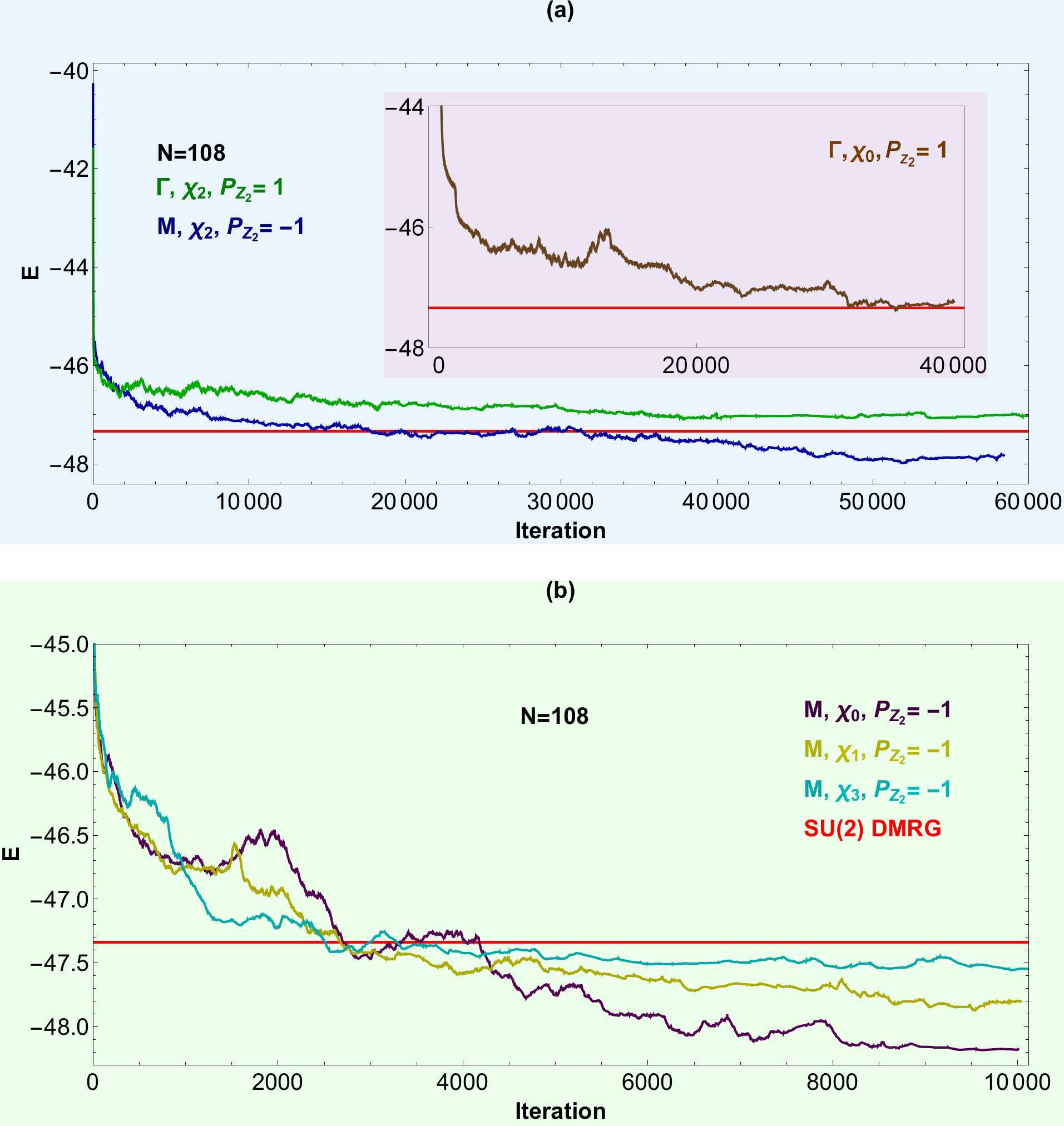}
\caption{\label{Fig:E_6x6x3}
Low energy eigenstates for the kagome lattice cluster with 108 sites ($6\times6\times3$ cluster) and periodic boundary conditions. The GCNN ans\"atze for the eigenstates have 4 layers and 4 features in each layer. Fully symmetric state (brown line graph in FIG. (a)) is obtained with SR minimization of the free energy loss defined in Eq. (\ref{eq:Free_energy}) and other eigenstates in FIG. (a) and FIG. (b) with SR minimization of the energy loss function (\ref{eq:Energy_VMC}) and with help of the transfer learning scheme described in the main text. The red line denotes SU(2) symmetric DMRG result for the ground state energy ($E_0^{DMRG}=-47.33964$).\cite{Depenbrock} Contrary to the result for the 48 lattice sites cluster, for larger 108 lattice sites cluster the ground state has nonzero crystal momentum corresponding to the $M$ high symmetry point in the kagome lattice Brillouin zone. Converged ground state energy value obtained with SR minimization (purple line graph in FIG. (b)), $E^{GCNN}=-48.18(0)$, has significantly lower energy than the ground state energy obtained with SU(2) symmetric DMRG calculations. Obtained values for the higher energy states are  $47.01(7)$, $-47.29(3)$, $-47.54(4)$, $-47.81(8)$ and $-47.85(5)$ (green, brown, cyan, yellow and blue line graphs, respectively).
}
\end{figure}
For the symmetry sector containing the lowest energy eigenstate with the total $S=1$ the algorithm however finds a low energy local minimum corresponding to a total $S=0$ eigenstate with two triplon ($S=1$) excitations forming a singlet. If triplons are formed by pairs of spins at sites $(i,j)$ and $(k,l)$, a spin parity $P_{\mathcal{Z}_2}=-1$ wavefunction will contain terms of the form  $\propto(|\uparrow_i\uparrow_j\downarrow_k\downarrow_l\rangle -|\downarrow_i\downarrow_j\uparrow_k\uparrow_l\rangle \bigotimes |\psi\rangle_{singlets} $. The nature of the found local minima can be seen by calculating the energy gap $\Delta=E_{\chi_1^0}^{GCNN}-E_{\chi_0^0}^{GCNN}\approx 0.36\pm 0.01$. It is approximately twice the triplet energy gap obtained by ED calculations $\Delta_t^{ED}=0.168217$. $\Delta_t^{GCNN}$ value however is $4-10\%$ higher value than $\Delta_t^{ED}$, which reflects either a more complicated sign structure of the higher energy states or presence of a weak repulsive interaction between the quasiparticles. We also note that for the random initial coefficients of the GCNN ans\"atze for the higher energy states the SR algorithm either becomes unstable or finds a local minimum with significantly higher energy.  As it will be demonstrated further, presence of such strong local minimum reflects instability towards spinon Cooper pair creation in the same symmetry sector for larger system sizes where described eigenstate becomes global minimum within the symmetry sector.  

To obtain a better estimate for the triplet energy gap we have also considered the total $S^z=-1$ sector. For $S^z_{tot}=-1$ the algorithm finds a local minimum that corresponds to three triplon excitations and provides a better estimate for the triplet gap $\Delta_t=0.17\pm 0.003$. 

We repeat the same procedure to find low-lying energy states for the kagome lattice cluster with 108 lattice sites ($6\times6\times3$ cluster) and periodic boundary conditions. These constitute the central result of our calculations. Whilst the ground state for the smaller lattice size is fully symmetric, we find that the ground state for larger cluster with 108 sites is in a nontrivial symmetry sector with nonzero lattice momentum revealing its long-range entanglement,\cite{Gioia} presence of Dirac points in the energy spectrum and instability towards spinon Cooper pair formation. We note that previous DMRG studies also found indications of the possible spinon Cooper pairing instability. In particular DMRG calculations\cite{Yan} found that the triplet excitations appear to be composed of two spinons. However they could not resolve whether the two spinons bind or not on cylindrical geometries.

Since increased number of symmetries for larger system sizes results in more GCNN parameters for $6\times6\times3$ cluster, GCNN ans\"atze with only 4 layers and 4 features in each layer already yield very accurate results. The SR minimization results for low energy eigenstates are shown in FIG. \ref{Fig:E_6x6x3}. The red line in FIG. \ref{Fig:E_6x6x3} denotes SU(2) symmetric DMRG result for the ground state energy.\cite{Depenbrock} After finding fully symmetric state (brown line graph in FIG. \ref{Fig:E_6x6x3}) with the lowest energy we have further checked, using again transfer learning approach, if any of the lowest energy states in other symmetry sectors for the crystal momentum corresponding to the $M$ high symmetry point (symmetry sectors shown in FIG. \ref{Fig:Characters_D6}) (b)) have even lower energy. Our calculations found the lowest energy for the eigenstate with characters $\chi_0$ in FIG. \ref{Fig:Characters_D6} (b), parity $P_{\mathbb{Z}_2}=-1$ and $\vec{q}=(\pi,\pi/ \sqrt 3)$ (purple line graph in FIG. \ref{Fig:E_6x6x3}) The obtained ground state energy $E_0=-48.18(0)$ is $\approx 1.78 \%$ lower than the ground state energy obtained with  DMRG calculations ($E_0^{DMRG}=-47.33964$).\cite{Depenbrock} We have also verified that the eigenstates for other crystal momenta and $P=-1$ have higher energy values than the eigenstates obtained for the crystal momentum corresponding to $M$ high symmetry point indicating that the spinon Cooper pairs have center of mass momentum $\vec{q}=\left(\pi,\pi/\sqrt3\right)$.

The results clearly demonstrate change in the symmetry sector of the ground state. We propose this is due to the spinon Cooper pairing instability close to the spinon Fermi surface, here being two Dirac points. The observed instability has previously been predicted as one of the possible instabilities of the "parent" U(1) Dirac spin liquid.\cite{Barkeshli} The instability is marginally relevant and it can be readily explained within fermionic parton theory.\cite{Wen,Lu1,Lu2} All other possible instabilities\cite{Barkeshli} would result in a ground state with zero crystal momentum, and this allows a PDW identification of the found ground state. 
Explanation of the Cooper pairing instability and formation of the found PDW within fermionic parton theory is presented in the next section. 

\begin{figure}[t!]
\includegraphics[width=\columnwidth]{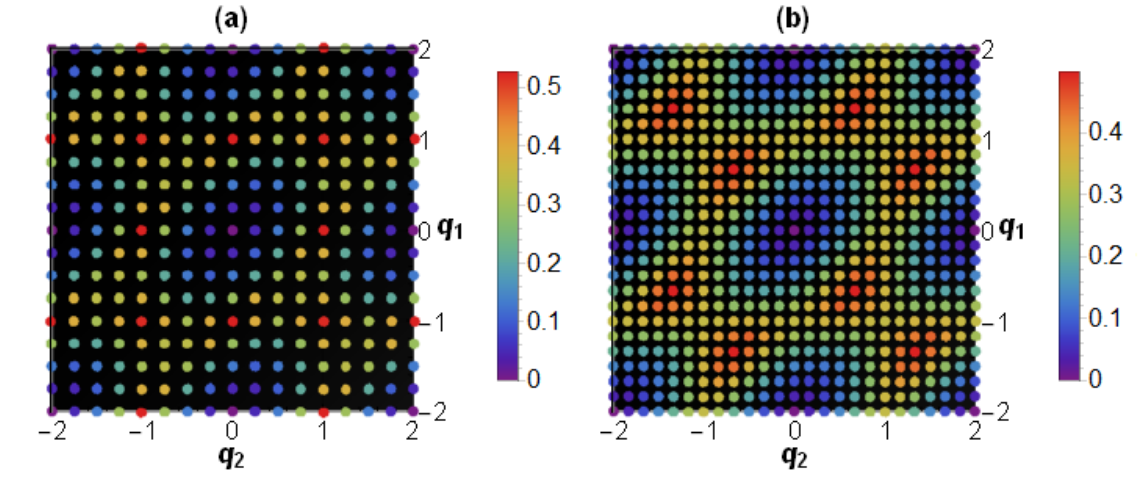}
\caption{\label{Fig:Spin_stucture_factor}The static spin structure factor $S_f^z(\vec{q})$ defined in Eq. (\ref{eq:spin_structure_factor}) for the system sizes of (a) 48 ($4\times 4\times 3$ cluster) and (b) 108 ($6\times 6\times 3$ cluster) lattice sites and with periodic boundary conditions calculated using optimized ground state GCNN ans\"atze. The structure factor is shown as a function of the wavevector $\vec{q}=\left(q_1,q_2\right)$ where components $q_1=n_1/N_1$ and $q_2=n_2/N_2$ are in units of the primitive basis vectors $\vec{b}_1$ and $\vec{b}_2$ shown in FIG. \ref{Fig:k_lattice} and $N_1=N_2=4$ in FIG. (a) and $N_1=N_2=6$ in FIG. (b).
}
\end{figure}
We have calculated spin and dimer structure factors for the found ground states to find additional signatures of the gapless spinon spectrum and spinon pair density wave order. The static spin structure factor
\begin{equation}\label{eq:spin_structure_factor}
S^z_f(\vec{q}\;)=\frac{1}{N}\sum_{ij}e^{i\vec{q}\left(\vec{r}_i-\vec{r}_j\right)}\langle S_i^zS_j^z\rangle
\end{equation}
is calculated using optimized ground state GCNN ans\"atze for both cluster sizes. We note that due to spin rotation symmetry $S_f^z(\vec{q}\;)=S_f(\vec{q}\;)/3$ with $S_f\left(\vec{q}\;\right)=\sum_{i\in \left\{x,y,z\right\}}S_f^i\left(\vec{q}\;\right)$; this also holds approximately when $SU(2)$ symmetry is only weakly broken. The wavevector $\vec{q}=\left(q_1,q_2\right)$ components $q_1=n_1/N_1$ and $q_2=n_2/N_2$ are in units of two primitive basis vectors $\vec{b}_1$ and $\vec{b}_2$ shown in FIG. \ref{Fig:k_lattice} with $n_i=\left\{0,...,N_i-1\right\}$ for $i\in{1,2}$ and $3\times N_1\times N_2=N$ being the number of lattice sites. The results are shown in FIG. \ref{Fig:Spin_stucture_factor} for $N=48$ and $N=108$ sites clusters. We note that since the unit cell contains three atoms the structure factor is  is not invariant under reciprocal lattice vector shifts. To fully describe correlations between all spins we calculate the structure factor in the extended Brillouin zone with $0\leq q_1\leq 2$ and $0\leq q_2\leq 2$. At large $|\vec{q}|$ the spectral weight is concentrated around the edge of the extended Brillouin zone with broad response and not very sharp or pronounced maxima, corresponding to the results obtained with DMRG\cite{Depenbrock} and projected fermionic wave-function\cite{Iqbal5} approaches. Whilst for the 48 sites cluster maxima are located at $(q_1,q_2)=(0,1)$, $(1,0)$ and $(1,1)$, for the 108 sites cluster maxima are found close to the M-points of the extended Brillouin zone and the structure factor has pinch-point like features that signal algebraically decaying spin-spin correlations in real space\cite{Kiese} and gapless spinon spectrum.

The results for the static spin structure factor clearly demonstrate that the ground state is a disordered state. The maxima of the structure factor do not scale with system size; they remain at approximately the same small value when the system size is increased ($\approx 0.526$ and $\approx0.498$ for 48 and 108 sites clusters, respectively) which is in sharp contrast to the behavior of the structure factor maxima for magnetic ordered systems. For the magnetic ordered phase the structure factor maximum becomes sharper and approaches delta-function Bragg peak with the system size increase whilst for disordered phases the structure factor maxima are broad and do not become sharper and higher with increasing system size.

The spin structure factor, however, does not clearly indicate the nature of the disordered ground state, since similar structure factor results are found for both $\mathbb{Z}_2$ and U(1) Dirac spin liquid phases.\cite{Depenbrock,Iqbal5} It also does not provide information about possible valence bond crystal (VBC) phases. We therefore also calculate dimer structure factor 
\begin{equation}\label{eq:dimer_structure_factor}
S_f^D(\vec{q})=\frac{1}{N}\sum_{i,j,k,l}e^{i\vec{q}\cdot\left(\vec{r}_{ik}-\vec{r}_{jl}\right)}D\left((i,k),(j,l)\right),
\end{equation}
where $\vec{r}_{ab}=(\vec{r}_a+\vec{r}_b)/2$ denote midpoints on each bond and 
\begin{equation}\label{eq:dimer-dimer_cf}
D((i,k),(j,l))=\langle \vec{S}_i\cdot\vec{S}_{k} \vec{S}_j\cdot\vec{S}_{l}\rangle-\langle \vec{S}_i\cdot\vec{S}_{k}\rangle \langle \vec{S}_j\cdot\vec{S}_{l}\rangle
\end{equation}
is the connected dimer-dimer correlation function with $k$ and $l$ being the nearest neighboring sites of $i$ and $j$.

\begin{figure}[t!]
\includegraphics[width=\columnwidth]{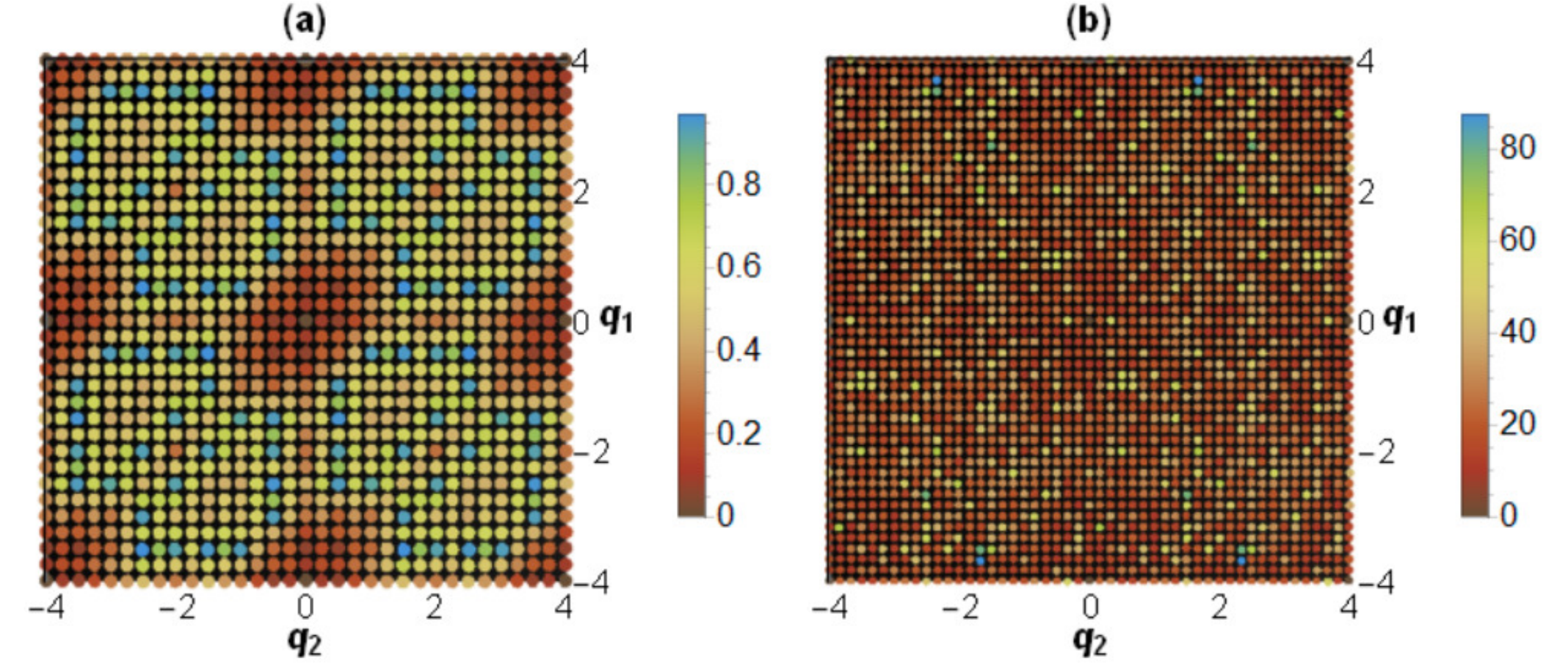}
\caption{\label{Fig:Dimer_sf}Dimer structure factor $S_f^z(\vec{q})$ defined in Eq. (\ref{eq:dimer_structure_factor}) for the system sizes of (a) 48 ($4\times 4\times 3$ cluster) and (b) 108 ($6\times 6\times 3$ cluster) lattice sites and with periodic boundary conditions calculated using optimized ground state GCNN ans\"atze. The structure factor is shown as a function of the wavevector $\vec{q}=\left(q_1,q_2\right)$ where components $q_1=n_1/N_1$ and $q_2=n_2/N_2$ are in units of the primitive basis vectors $\vec{b}_1$ and $\vec{b}_2$ shown in FIG. \ref{Fig:k_lattice} and $N_1=N_2=4$ in FIG. (a) and $N_1=N_2=6$ in FIG. (b).
}
\end{figure}

The results for the dimer-dimer structure factor for 48 and 108 lattice sites clusters are shown in FIG. \ref{Fig:Dimer_sf}. For the 48 sites cluster the structure factor $S^D\left(\vec{q}\right)$ does not have very sharp peaks. The results are similar to the results obtained for the same cluster with DMRG calculations\cite{Jiang} with slightly higher maxima values ( maximum value $\approx 0.933$). However, contrary to the DMRG results, the overall magnitude of the dimer structure factor significantly increases for the 108 sites cluster. The structure factor exhibits numerous peaks, with several more pronounced peaks and its maximum value is $\approx 87.370$. This behavior reflects the existence of the spinon PDW and the secondary order parameters induced by the spinon PDW. Since the instability towards finite momentum Cooper pairs leads to PDW ground state only for larger system sizes, to further study dimer structure factor scaling with the system size, further GCNN and VMC calculations for larger system sizes are necessary. This computationally very challenging task is beyond the scope of our current study and is one of our future research directions. 

Before proceeding to the explanation of the found state in the following section we briefly comment on the accuracy of the obtained results, quality of the obtained wavefunctions and the sampling scheme for both cluster sizes. In general for the systems with large number of possible competing many-body states, as it is the case for kagome lattice antiferromagnet, even small relative errors of order $10^{-3}$ in variational energy may result in the optimized ansatz wavefunction with correlations significantly different from the correlations of the true ground state. Presence of large number of the competing ground states is reflected in large number of low-lying total spin $S=0$ singlet excitations below the first total spin $S=1$ triplet excitation in the energy spectrum\cite{Lauchli} that can host such possible competing states. A relative error of order $10^{-3}$ corresponds to several times singlet energy gap for smaller system sizes.\cite{Lauchli} However, within GCNN approach the low-lying singlet states correspond to different space-group irreducible representations with different space group characters $\chi_g$ and a small relative error for a chosen irreducible representation reflects a small quantitative rather than qualitative error that would result in significantly different ground-state wavefunction correlations. We also point out that the ground state energies obtained within presented GCNN approach provide significantly more accurate results than approaches based on other neural network architectures\cite{Westerhout,Fu,Kochkov}, reflecting the importance of introducing symmetry equivariance in neural network architectures.

We also note that indication of the spinon Cooper pairing instability and change of the ground state symmetry sector for the larger cluster can already be seen from the sampling schemes that yield the best ground state energies. Whilst for the 48 sites cluster the best variational energy is obtained by the free energy minimization within the total $S^z=0$ sector where the samples are generated by exchanging spins at two randomly chosen lattice sites, for the 108 sites cluster the best results are found by sampling within the whole Hilbert space with samples generated by flipping a spin at random lattice site. The ground state wave function quality can be estimated by evaluating $\langle \vec{S}^2 \rangle$ which ideally should be zero since the exact ground state wavefunction has SU(2) symmetry. 

For the 108 sites cluster we find $\langle \vec{S}^2 \rangle\approx 2.5\times 10^{-7}$ confirming very high quality of the obtained ground state wave function. For the 48 sites cluster we however find somewhat larger value of $\langle \vec{S}^2 \rangle\approx 0.14$. The obtained value still indicates good quality of the obtained ground state wave function with only weakly broken SU(2) symmetry.\cite{Roth} We also note that minimization within the whole Hilbert space for the 48 sites cluster, with samples generated by flipping a spin at random lattice site, yields better result for $\vec{S}^2$ ($\approx 10^{-7}$) and less accurate result for the ground state energy. Moreover, such minimization requires larger number of samples, exhibits slower convergence and finds significantly higher energy values for the low-lying energy states with more complicated sign structures than the energies found by minimization within the total $S^z=0$ sector that are shown in FIG. \ref{Fig:E_4x4x3}.

\section{Spinon pair density wave ground state description within fermionic parton theory}
\label{sec:PartonConstruction}
In the Abrikosov fermion parton construction a spin-1/2 operator at site $i$ is represented by fermionic spinons $f_{i\alpha}$ with $\alpha \in{\uparrow,\downarrow}$:
\begin{equation}\label{eq:spinons}
\vec{S}_i=\frac{1}{2}f_{i\alpha}^{\dagger}\vec{\sigma}_{\alpha\beta}f_{i\beta},
\end{equation}
where $\sigma=(\sigma^x,\sigma^y,\sigma^z)$ are Pauli matrices. Within mean-field theory symmetric spin liquids are obtained by introducing mean-field parameters $\chi_{ij}$ and $\Delta_{ij}$:
\begin{eqnarray}\label{eq:mf_parameters}
\chi_{ij}\delta_{\alpha\beta}=2\langle f_{i\alpha}^{\dagger} f_{j\beta}\rangle,\\
\Delta_{ij}\epsilon_{\alpha\beta}=-2\langle f_{i\alpha}f_{j\beta}\rangle, \nonumber
\end{eqnarray}
where $\delta_{\alpha\beta}$ is Kronecker delta and $\epsilon_{\alpha,\beta}$ is the completely antisymmetric tensor. Since the parton construction enlarges the Hilbert space of the original spin model, the physical spin state is obtained by introducing constraint of one spinon per site:
\begin{eqnarray}\label{eq:mf_constraint}
f_{i\alpha}^\dagger f_{i\alpha}=1,\\
f_{i\alpha}f_{i\beta}\epsilon_{\alpha\beta}=0,\nonumber
\end{eqnarray}
where repeated indices are summed over. Within the parton theory the magnetic excitations then correspond to deconfined spinons and non-magnetic (total spin $S=0$) excitations to fluxes (or vortices) of the $Z_2$ gauge field called visons.

\begin{figure}[t!]
\includegraphics[width=\columnwidth]{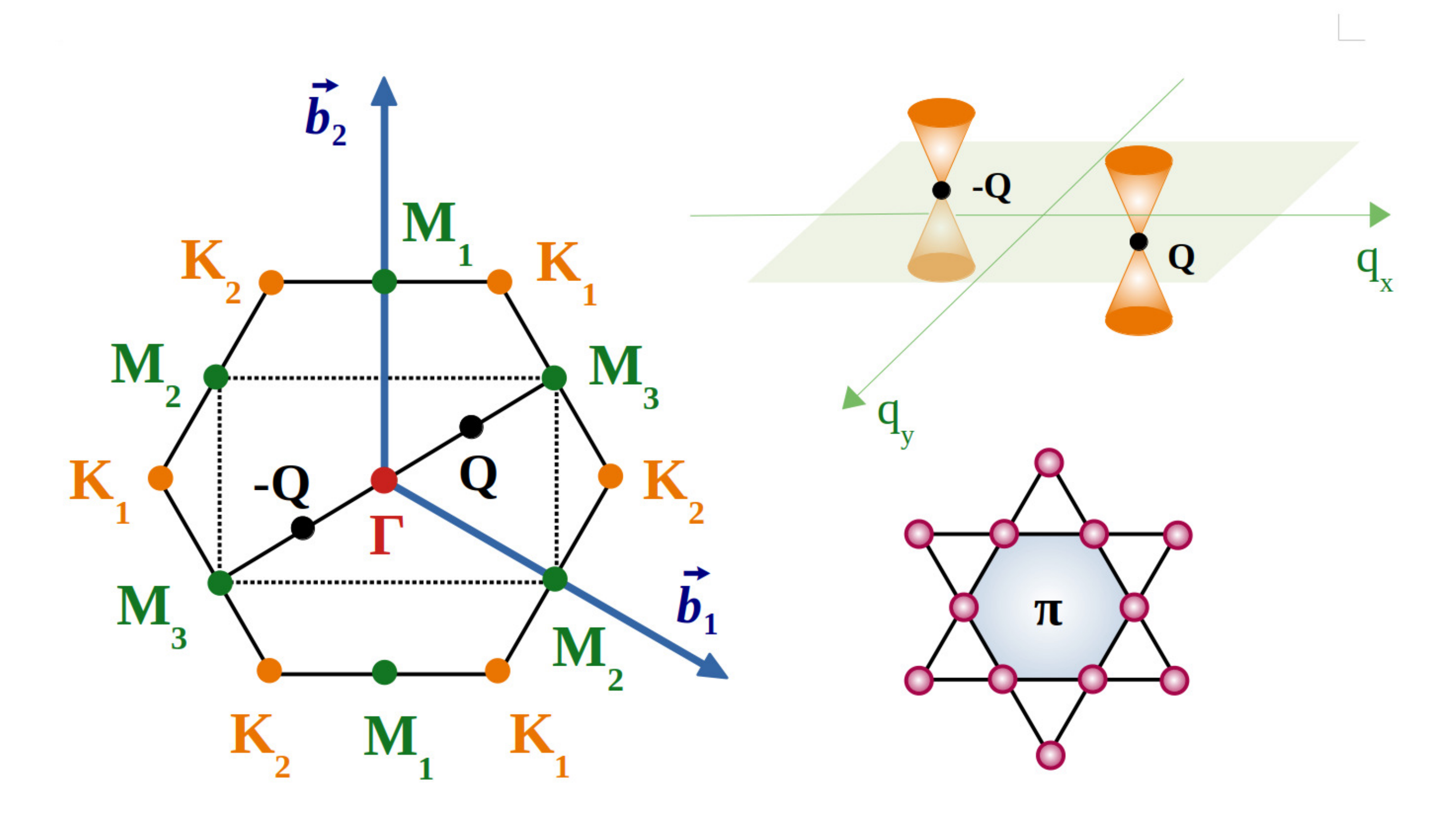}
\caption{\label{Fig:Dirac_spin_liquid}Two Dirac nodes of the Dirac spin liquid (U(1) SL-$\left[0,\pi\right]$ state), denoted here by $\pm\vec{Q}$. Magnetic Brillouin zone of the Dirac spin liquid is halved (dashed lines) since it is a $\pi$-flux ansatz with zero flux through the triangles and $\pi$-flux through the hexagons of the kagome lattice. 
}
 \end{figure}

Projective symmetry group (PSG) calculations with fermionic partons find both gapped and gapless symmetry allowed $Z_2$ spin liquid states.\cite{Lu1,Lu2}These spin liquid states are continuously connected to different parent $U(1)$ gapless spin liquid states that in general have the following mean-field ansatz
\begin{equation}\label{U1_SL}
H_{U(1)SL}=\chi\sum_{\langle ij\rangle}\nu_{ij}(f_{i\alpha}^\dagger f_{j\alpha}+h.c.),
\end{equation}
where $\chi\in\mathbb{R}$ and $\nu_{ij}=\pm 1$ characterizes signs of nearest-neighboring hopping terms. Different $U(1)$ spin liquids then correspond to different spinon hopping phases around hexagonal and triangular plaquettes of the kagome lattice. VMC calculations showed that the U(1) Dirac spin liquid state with zero-flux through the triangles and $\pi$-flux through the hexagons (U(1) SL-$\left[0,\pi\right]$ state) has considerably lower energy compared to many other competing states.\cite{He,Ran,Iqbal2,Iqbal3,Iqbal4,Iqbal5,Hastings,Iqbal6, Iqbal7} VMC calculations have also shown that all gapped and gapless $Z_2$ spin liquids are higher in energy compared to the parent $U(1)$ gapless spin liquids those $Z_2$ spin liquids are continuously connected to.\cite{Iqbal2} The magnetic Brillouin zone and positions of two Dirac nodes of the $U(1)$ Dirac state are illustrated in FIG. \ref{Fig:Dirac_spin_liquid}. Such state in two-dimensions has an emergent SU(4) symmetry.

\begin{figure}[b!]
\includegraphics[width=\columnwidth]{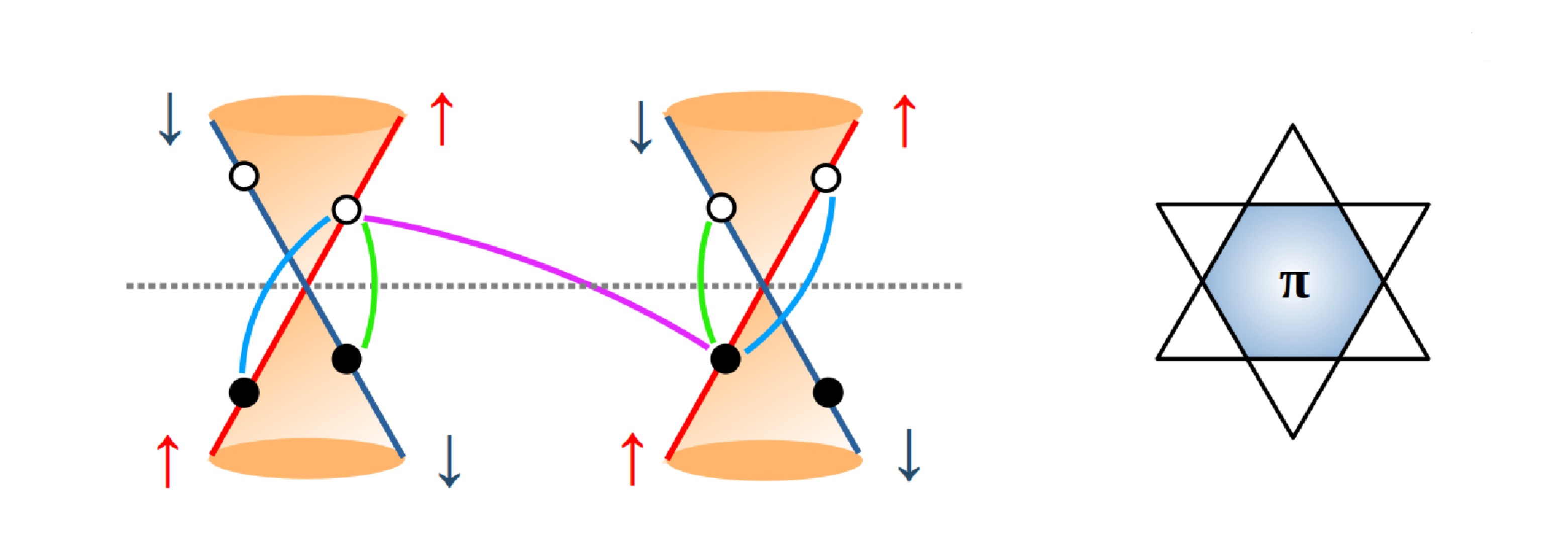}
\caption{\label{Fig:partice-hole_excitations} The lowest energy inter-valley and intra-valley particle-hole excitations in the vicinity of the Dirac points. 
}
\end{figure}

Within fermionic parton theory spin triplet (total $S=1$) excitations correspond to the particle-hole excitations near Dirac points, since a spin flip with momentum $\vec{q}_1-\vec{q}_2$ can be written as $S^{+}(\vec{q}_1-\vec{q}_2)=f_{\uparrow}^{\dagger}(\vec{q}_1)f_{\downarrow}(\vec{q}_2)$. The lowest energy inter-valley and intra-valley particle-hole excitations are illustrated in FIG. \ref{Fig:partice-hole_excitations}. DMRG calculations of the transfer matrix spectrum in the $S^z=1$ sector\cite{He} found Dirac cone structure for the particle-hole mode denoted by magenta line in FIG. \ref{Fig:partice-hole_excitations} indicating the presence of Dirac points in the spectrum of the kagome lattice antiferromagnetic Heisenberg model. Also, for the gapless $Z_2$ spin liquid states found within PSG approach\cite{Lu1,Lu2} Dirac nodes in the spinon spectrum are protected by symmetry. 

Our results however show that for the highly symmetric $6\times6\times3$ cluster that contains all three high symmetry points in the Brillouin zone ($\Gamma$, M and K points in FIG. \ref{Fig:k_lattice}) the ground state has \emph{finite crystal momentum}. We further argue that the found state is a spinon-pair density wave (PDW) state that does not break time-reversal symmetry nor any of the lattice symmetries. This is supported by the results for the chiral order parameter and spin and dimer structure factors presented further in this section.

PDW state is a state in which Cooper pairs (in this case spinon Cooper pairs) have ﬁnite center of mass momentum. The first found example of PDW state is the Fulde–Ferrell–Larkin–Ovchinnikov (FFLO) state\cite{Fulde,Larkin} which appears in superconductors in the presence of a Zeeman magnetic field and breaks time-reversal symmetry. It was subsequently conjectured that for systems with strong interactions a PDW could occur without time-reversal symmetry breaking and was studied in the context of underdoped cuprate superconductors in connection with pseudogap phase.\cite{Agterberg,Lee3,Lee4,Xu,Chakraborty} For such PDW state the spatial average of the superconducting order vanishes which allows the state to be distinguished from the coexisting  charge density wave and superconducting order.

In a PDW state the mean-field parameter that represents Cooper pairs  
\begin{equation}\label{eq:Cooper_pairs}
\Delta^*_{\sigma,\sigma'}(\vec{r}_i,\vec{r}_j)\equiv \langle f_{\sigma}^\dagger(\vec{r}_i)f_{\sigma'}^\dagger(\vec{r}_j)\rangle
\end{equation}
is not translation invariant, that is $\Delta^*(\vec{r}_i,\vec{r}_j)\neq \Delta^*(\vec{r}_i+\vec{R},\vec{r}_j+\vec{R})$, where $\vec{R}$ is any Bravais lattice vector. In the context of spin liquids that would mean that $\Delta^*_{\sigma,\sigma'}(\vec{r}_i,\vec{r}_j)$ is not invariant with respect to combined lattice translation and gauge group transformations within the PSG approach. Symmetries of the translation invariant states can be classified by the point group irreducible representations (for example s-wave, d-wave or p-wave states), properties under spin rotations in the absence of spin-orbit coupling (singlet or triplet states), and according to the behavior with respect to time reversal where in general the state can be time-reversal symmetric or break the symmetry (for example $p_x+ip_y$ state). For spinons corresponding states are $Z_2$ spin liquids obtained within PSG approach. 

Contrary to translation invariant states, a PDW state exhibits spatial modulation, that has nontrivial dependence on $\vec{R}$. Whilst $\Delta^*(\vec{r}_i,\vec{r}_j)$ is nonzero for a PDW, its spatial average 
\begin{equation}\label{eq:Cooper_pairs_average}
\Delta^{*\left(0\right)}_{\sigma,\sigma'}(\vec{r}_i,\vec{r}_j)=\frac{1}{N_{u.c.}} \sum_{\vec{R}} \Delta^{*}_{\sigma,\sigma'}(\vec{r}_i+\vec{R},\vec{r}_j+\vec{R})=0
\end{equation}
where $N_{u.c.}$ is the number of unit cells, and a PDW state is characterized by the smallest value of the wavevector $\vec{P}$ for which 
\begin{equation}\label{eq:Cooper_pairs_average_2}
\Delta^{*(\vec{P}\;)}_{\sigma,\sigma'}(\vec{r}_i,\vec{r}_j)=\frac{1}{N_{u.c.}} \sum_{\vec{R}} e^{i\vec{P}\vec{R}}\Delta^{*}_{\sigma,\sigma'}(\vec{r}_i+\vec{R},\vec{r}_j+\vec{R})
\end{equation}
has non-zero value. Similar average will then also be nonzero at wavevectors $n\vec{P}$.

If Cooper pairs are formed between fermions (spinons) with momenta $\vec{Q}-\vec{k}$ and $\vec{Q}+\vec{k}$ near point $\vec{Q}$ on the Fermi surface, the pairs carry momenta $\vec{P}=2\vec{Q}$. Once such pairing occurs, $U(1)$ gauge group is reduced to $Z_2$. As for nodal $Z_2$ spin liquids (of which the Kitaev spin liquid is a known example\cite{Kitaev}) the spinon Fermi surface of the PDW is not fully gapped; instead, it features gapless Dirac-like nodal points in the spinon spectrum. This is because the gap parameter changes sign when translated for 1/2 of the PDW period. The Dirac-like nodal points for the nodal spin liquids are known to be perturbatively stable as long as time-reversal symmetry is not broken.\cite{Berg2} We further point out that although a PDW in general breaks lattice translation symmetry, for the special case when PDW period is two lattice constants the lattice translation symmetry is not broken, since translation for one lattice constant corresponds to a uniform gauge transformation. In that case the PDW state can occur without breaking any lattice symmetries. In addition if the Cooper pairs with momenta $\vec{P}$ and $-\vec{P}$ coexist, the corresponding two component mean-field parameter also does not break time reversal symmetry\cite{Loder} and has the form
\begin{equation}\label{eq:Order_parameter}
\Delta^*_{\sigma,\sigma'}(\vec{r}_i,\vec{r}_j)=F(\vec{r}-\vec{r}')[\;\Delta^{*(\vec{P}\;)}_{\sigma,\sigma'}\cdot e^{i\vec{P}\vec{R}}+\Delta^{*(-\vec{P}\;)}_{\sigma,\sigma'}\cdot e^{-i\vec{P}\vec{R}}\;],
\end{equation}
where $F(\vec{r}-\vec{r}')$ is a short-range function (for example d-wave like) and $\vec{R}=(\vec{r}_1+\vec{r}_2)/2$.

For the obtained ground state of the 108 lattice sites kagome cluster, the Fermi surface corresponds to the two Dirac nodes illustrated in FIG. \ref{Fig:Dirac_spin_liquid}. The spinon Cooper pairs are formed from spinons with momenta $\pm\vec{Q}-\vec{k}$ and $\pm\vec{Q}+\vec{k}$ resulting in a PDW with $\vec{P}=2\vec{Q}={(\vec{b}_1+\vec{b_2})/2}$. Here $\vec{b}_1$ and $\vec{b}_2$ are two primitive basis vectors in the reciprocal lattice (FIG. \ref{Fig:k_lattice}) and $\vec{P}$ corresponds to the M high symmetry point in the Brillouin zone. Therefore the PDW has period of two lattice constants and does not break translation symmetry or any of the point group symmetries. The PDW mean-field parameter coupling to spinons can be described with 
\begin{equation}\label{eq:PDW_coupling}
H_{\vec{P}}^{\sigma\sigma'}=\Delta_{\vec{P}}^{\sigma\sigma'}\sum_{\vec{k}}F(\vec{k})f_{\sigma,\vec{k}+\frac{\vec{P}}{2}}^{\dagger}f_{\sigma',-\vec{k}+\frac{\vec{P}}{2}}^{\dagger},
\end{equation}
where $F(\vec{k})$ is an internal form factor that can be, for example, $s$-wave or $d$-wave or mixed. For the PDW with $\vec{P}/2=\vec{Q}$ and $d$-form factor Dirac points in the spinon spectrum remain gapless. The gauge fluctuations are gapped due to reduction of $U(1)$ gauge group to $Z_2$ as it can be seen in FIG. \ref{Fig:E_6x6x3} (the state represented by green line that corresponds to visons that are vortex-like excitations). Furthermore the points $\vec{P}$ and $-\vec{P}$ are equivalent since they are connected by a reciprocal lattice vector and the state does not break time reversal symmetry. We confirm this by calculating the chiral order parameter\cite{Gong}
\begin{equation}\label{eq:chiral_order_parameter}
\chi_i=\langle\vec{S}_{i_1}\cdot(\vec{S}_{i_2}\times \vec{S}_{i_3})\rangle,
\end{equation}
where $i$ denotes the triangular plaquettes of the kagome lattice ($i\in \triangle_i,\bigtriangledown_i$). The chiral order parameter vanishes for the ground states obtained for both 108 and 48 lattice sites clusters. 

We further note that a PDW can in general induce secondary orders. Studying such induced orders has great importance in understanding the relevance of the PDW orders in cuprates, \cite{Lee3,Fradkin,Berg,Agterberg2,Berg3,Agterberg3,Wang2,Dai} primarily charge density wave (CDW) orders and also Ising nematic, magnetization density wave (MDW) and translation invariant charge-$4e$ superconducting orders. For the found ground state, the possible induced order parameters include CDW, nematic and 4 spinon order parameters, where 4 spinon order parameter is equivalent to a charge-$4e$ superconducting order parameter for electrons. If Cooper pairs form at momenta $\vec{P}$ and $-\vec{P}$, CDW order appears as a second harmonic of the fundamental PDW order and has the form $\rho_{2\vec{P}}\propto \Delta_{\vec{P}}\Delta_{-\vec{P}}^{*}$ with ordering wavevector $\vec{K}=2\vec{P}$, whilst nematic ordering has the form $N\propto |\Delta_{\vec{P}}|^2 +|\Delta_{-\vec{P}}|^2$. In general CDW order breaks lattice translation symmetry whilst nematic order breaks lattice rotation symmetry. However for the PDW period equal to two lattice constants, as it is the case for the kagome lattice PDW where $\vec{P}$ corresponds to the M high symmetry point wavevector and $\vec{P}$ and $-\vec{P}$ differ by a reciprocal lattice vector, lattice translation and rotation symmetries are not broken. 

The 4 spinon order parameter is
\begin{equation}\label{eq:4e_order_parameter}
\Delta^{*(4)}(i,j,k,l)\equiv\langle f_{\sigma_i}^{\dagger}(\vec{r}_i)f_{\sigma_j}^{\dagger}(\vec{r}_j)f_{\sigma_k}^{\dagger}(\vec{r}_k)f_{\sigma_l}^{\dagger}(\vec{r}_l)\rangle,
\end{equation}
where $i\equiv(\vec{r}_i,\sigma_i)$.
If $\Delta^{*_{\sigma,\sigma'}}(\vec{r},\vec{r}')\neq 0$ then also some components of the $\Delta^{*(4)}(i,j,k,l)$ order parameter will be nonzero, which can be seen by rewriting the order parameter at the mean field level as
\begin{eqnarray}\label{Delta4_Delta2}
\Delta^{*(4)}(i,j,k,l)&\sim& \Delta^*(i,j)\Delta^*(k,l)+\Delta^*(i,l)\Delta^*(j,k) \nonumber\\
&-&\Delta^*(i,k)\Delta^*(j,l).
\end{eqnarray}
Although the uniform component $\Delta^{*\left(0\right)}$ described by Eq. (\ref{eq:Cooper_pairs_average}) vanishes, the uniform component of $\Delta^{(4)}$ which is of the form   $\Delta_{\vec{P}}\Delta_{-\vec{P}}$ does not vanish. Such ordering in some circumstances can be more robust than PDW ordering, as in cases when thermal or quantum fluctuations destroy PDW order however not charge-$4e$ order \cite{Berg4,Radzihovsky} whilst Fermi surface of gapless quasiparticles remains. In the context of spin liquids such state would be a $Z_4$ spin liquid with gapped gauge fluctuations and spinon Fermi surface.\cite{Barkeshli} A $Z_4$ spin liquid cannot be described within a simple mean-field theory that is quadratic in terms of fermion operators. Also vortex-like excitations (visons) of the $Z_4$ gauge field carry $1/4$ of the flux quantum, and since a $Z_4$ vison is different from its time-reversed anti-vison with the flux $-1/4$ of the flux quantum, visons in a $Z_4$ spin liquid necessarily break time-reversal symmetry. We have found that all eigenstates found within our GCNN and VMC approach have vanishing chiral order parameter confirming time-reversal symmetry of the eigenstates and indicating presence of the spinon PDW ordering.

\section{Conclusions}
\label{sec:Conclusions}
We have studied properties of the spin-1/2 kagome antiferromagnet using GCNN and VMC approach, a recently developed advanced machine learning technique that allows studying strongly frustrated models with high accuracy. GCNNs are deep neural network architectures equivariant with respect to all space group transformations that in addition to expressivity also possess ability to efficiently learn from a limited number of samples allowing efficient neural network training and more accurate solutions for the optimal network parameters. Our study therefore provides an important step in clarifying the kagome antiferromagnet properties and a reference to check validity of the results obtained with other methods. 

Contrary to the results obtained with various other methods, that predicted $Z_2$ or U(1) Dirac spin liquid states, we have found that the ground state of the kagome lattice antiferromagnet is a spinon PDW that does not break time-reversal symmetry or any of the lattice symmetries. The found ground state has significantly lower energy than the lowest energy states found by the SU(2) symmetric density matrix renormalization group calculations and other methods. The PDW state  appears due to the spinon Cooper pairing instability close to two Dirac points in the spinon energy spectrum resulting in formation of spinon Cooper pairs with ﬁnite center of mass momentum. Further examination of the found spinon Cooper pairing instability requires calculation of the system properties for larger system sizes. This computationally very challenging task is one of our future research directions together with the formulation of an effective theory that would capture PDW formation on the kagome lattice and possible PDW induced secondary orders. Additional studies are also necessary to examine the stability of such states to various perturbations, for example further neighbor couplings and behavior of the system upon doping studied previously assuming U(1) Dirac spin liquid ground for undoped kagome system.\cite{Ko} 

\begin{acknowledgments}
We thank Dhiman Bhowmick, Jon David Spalding, Kathyat Deepak Singh, Yasir Iqbal, Dario Poletti, Daniel Leykam, Wei Zhu, Shoushu Gong, Christopher Roth and Anders Sandvik for helpful discussions. BY acknowledges support by the National Research Foundation, Singapore under the NRF fellowship award (NRF-NRFF12-2020-005). This work was supported by the Google cloud research credits academic research grant, and Singapore Ministry of Education (MOE) Academic Research Fund Tier 3 Grant (No. MOE-MOET32023-0003) "Quantum Geometric Advantage". We would also like to acknowledge the NTU High Performance Computing Centre for providing computing resources, facilities, and services that have contributed to this work.
\end{acknowledgments}

\begin{appendix}
\section{Group convolutional neural networks}
\label{Appendix_GCNN}
\begin{figure}[!b]
\includegraphics[width=\columnwidth]{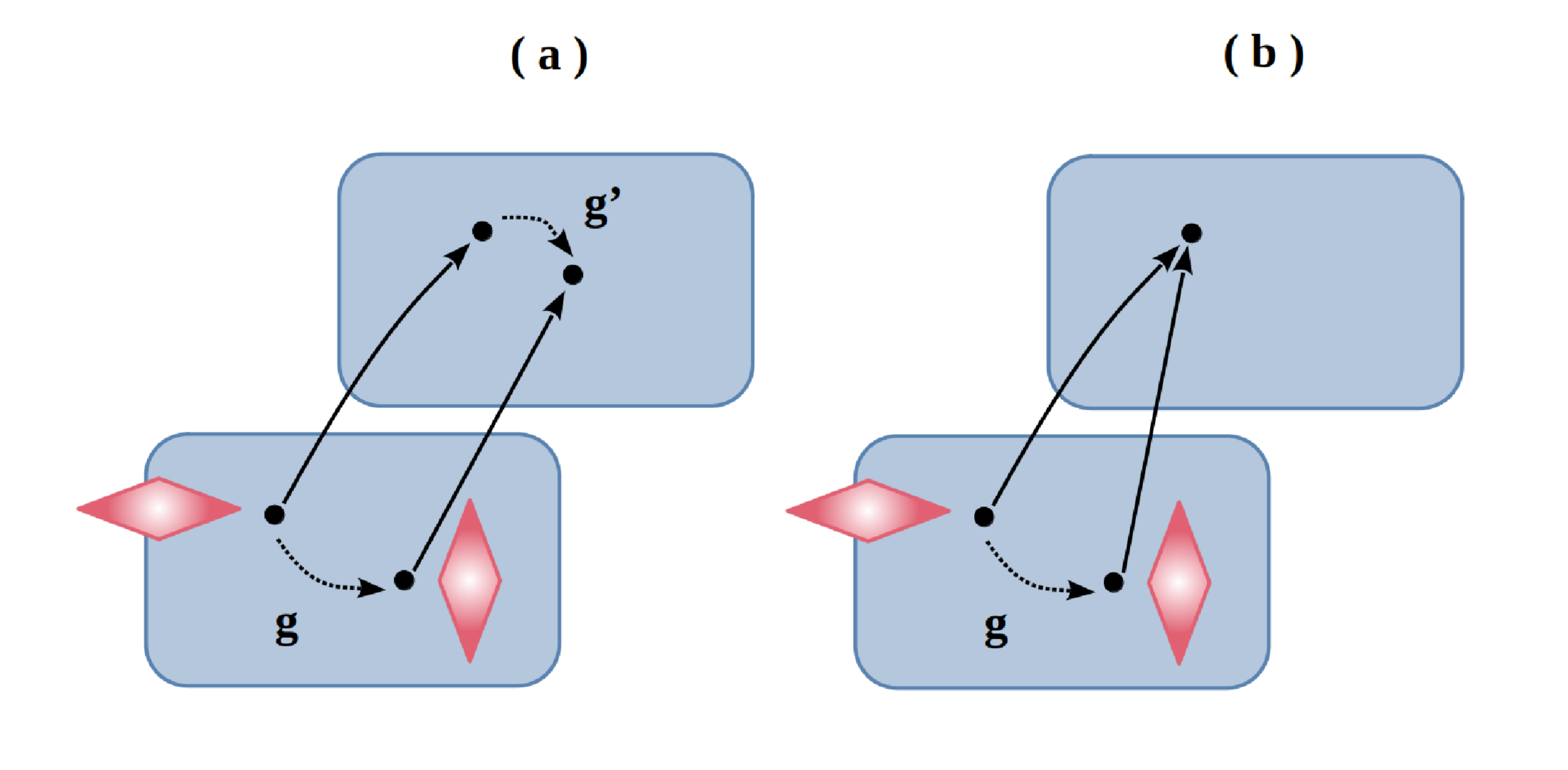}
\caption{\label{Fig:Equivariance} Schematic illustration of (a) equivariance: $f(gx)=g'f(x)$, and (b) invariance: $f(gx)=f(x)$. 
}
\end{figure}

With a symmetry group denoted by $G$, input denoted by $x$ and transformations that correspond to the group elements by $g$, a function, layer or network $f$ is equivariant with respect to the transformations $g$ if
\begin{equation}\label{eq:equivariance_a}
f(gx)=g'(f(x))
\end{equation}
for any $g$, where $g$ and $g'$ do not in general need to be the same. As illustrated in FIG. \ref{Fig:Equivariance} invariance is obtained if $g'$ is identity map for all transformations $g$. In other words, $f$ preserves the symmetry group structure in the sense that acting on inputs by symmetry transformations causes outputs to be transformed by the same symmetry elements in a non-trivial way. 

\begin{figure}[b!]
\includegraphics[width=\columnwidth]{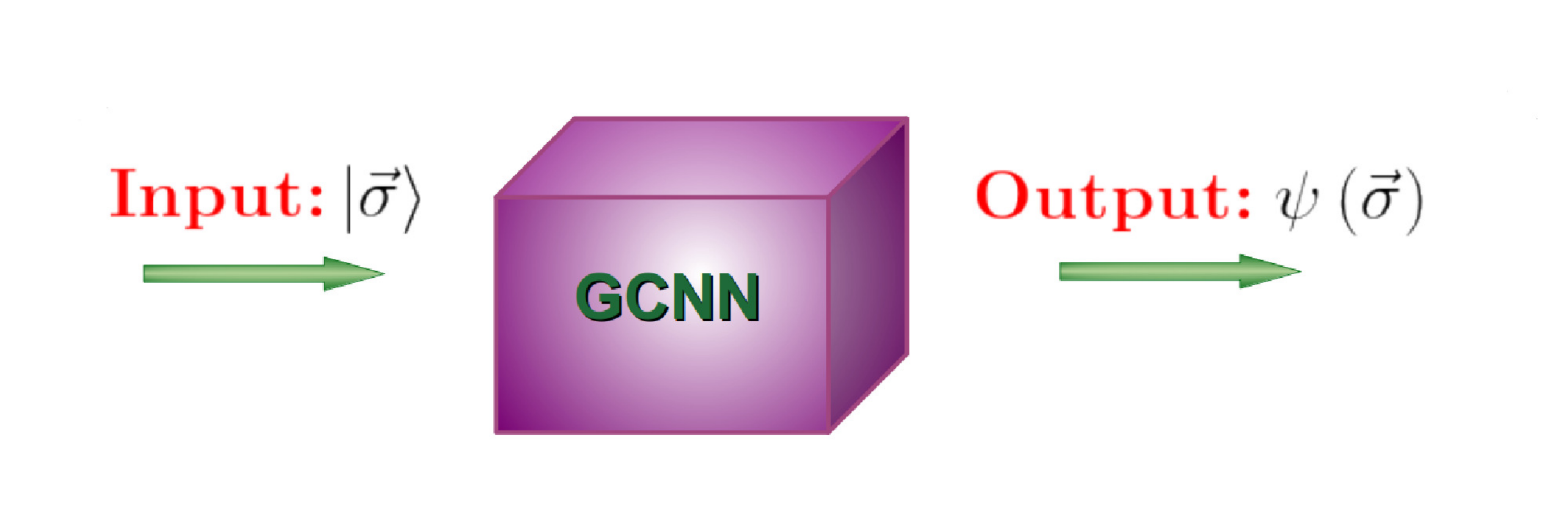}
\caption{\label{Fig:GCNN} Spin configurations $|\vec{\sigma}\rangle$ are GCNN inputs, and the complex ansatz wave-function coefficients $\psi(\vec{\sigma})$ GCNN outputs. 
\\~
}
\end{figure}

For systems of interacting spins on a lattice neural network ans\"atze for the wavefunctions $|\psi\rangle$ of the lattice Hamiltonian associate a complex number $\psi(\vec{\sigma};\vec{\alpha})$ with each computational basis configuration of spins $|\vec{\sigma}\rangle$ where $\vec{\alpha}$ denotes the network parameters:
\begin{equation}\label{eq:ansatz1_a}
|\psi\rangle=\sum_{\vec{\left\{\sigma\right\}}}\psi\left(\vec{\sigma};\vec{\alpha}\right)|\vec{\sigma}\rangle,
\end{equation}
and $\vec{\sigma}=\left\{\sigma(\vec{r}_1),...,\sigma(\vec{r}_N)\right\}$ with $\vec{r}_1,...,\vec{r}_N$ being positions of the lattice sites. Spin configurations $|\vec{\sigma}\rangle$ are the network inputs, and the complex coefficients $\psi(\vec{\sigma})$ are the network outputs as illustrated in FIG. \ref{Fig:GCNN}.

For example, for a square lattice with periodic boundary conditions a CNN convolutional mapping is
\begin{equation}\label{eq:CNN1_a}
\psi\left(\vec{r}\right)=\sum_{\vec{r\sp{\prime}}}K\left(\vec{r}-\vec{r\sp{\prime}}\right)\sigma\left(\vec{r\sp{\prime}}\right),
\end{equation}
where $K$ is an arbitrary kernel matrix that looks for and filters patterns in $\vec{\sigma}$.  The mapping is clearly translation equivariant since 
\begin{eqnarray}\label{eq:CNN2_a}
&&\sum_{\vec{r\sp{\prime}}}K\left(\vec{r}-\vec{r\sp{\prime}}\right)\sigma\left(\vec{r\sp{\prime}}+\vec{t} \;\right)=\sum_{\vec{r\sp{\prime}}}K\left(\vec{r}+\vec{t}-\vec{r\sp{\prime}}\right)\sigma\left(\vec{r\sp{\prime}}\right) \nonumber \\
&&=\psi\left(\vec{r}+\vec{t}\;\right). 
\end{eqnarray}
It can similarly be shown that several iterations of convolutional mappings of the form (\ref{eq:CNN1_a}) combined with arbitrary pointwise non-linear functions, and therefore neural networks built from layers of such mappings, are also translation equivariant. A translation for $\vec{t}$ acts by adding $\vec{t}$ to both input and output coordinates and shifting input data and then passing the data through a number of layers is the same as passing the original data through the same layers and then shifting the resulting output feature maps. 

In GCNNs convolutions of the form (\ref{eq:CNN1_a}) are generalized to arbitrary discrete nonabelian groups. For a symmetry group G with elements $\hat{g}$ an embedding layer first generates feature maps from the input data
\begin{equation}\label{eq:GCNN_K_a}
\mathbf{f}_g^1=\Gamma\left(\sum_{\vec{r}}\mathbf{K}\left(\;\hat{g}^{-1}\vec{r}\;\right)\sigma\left(\;\vec{r}\;\right)\right)
\end{equation}
for a set of learnable kernels or filters $\mathbf{K}$ that extract local features from the input and $\Gamma$ is a non-linear activation function. The output of the embedding layer is further passed to any number of group convolutional layers denoted by $i=2,..,N_l$ where feature to feature convolutions are performed as 
\begin{equation}\label{eq:GCNN_W_a}
\mathbf{f}_g^{i+1}=\Gamma\left(\sum_{\hat{h}\in G}\mathbf{W}^i\left(\hat{h}^{-1}\hat{g}\right)\mathbf{f}_h^i\right)
\end{equation}
for a set of convolutional kernels $\mathbf{W}$. Depth of the filters $\mathbf{K}$ is one (the same depth as input data) and depth of the filters $\mathbf{W}^i$ equals to the number of feature maps in the $i$-th layer. Here $\Gamma$ is taken to be scaled exponential linear unit (SELU) activation function:
\begin{equation}\label{eq:SELU_a}
\mbox{SELU}(x)=a
\begin{cases}
x\;\;\; x > 0 & \\
b(e^x-1 )\;\;\; x\leq 0
\end{cases}
\end{equation}
with $a\approx 1.05$ and $b\approx 1.67$, and applied separately to the real and imaginary components:
\begin{equation}\label{eq:Gamma__a}
\Gamma(x) = \mbox{SELU}(\mbox{Re}(x)) + i \mbox{SELU}(\mbox{Im}(x)).
\end{equation}

The SELU activation functions were introduced for self-normalizing neural networks\cite{Klambauer} since these activations generate self-normalizing properties, namely variance stabilization that helps in avoiding exploding and vanishing gradients when training deep neural networks, i.e. when optimizing variational wave function parameters. The SELU nonlinearity moves distribution of the nonlinearity inputs for the network layers towards mean being zero and variance one allowing stable training for very deep neural networks.

All $N_l$ hidden layers are followed by the $\Gamma$ activation function. To obtain $\psi(\sigma)$ in Eq. (\ref{eq:ansatz1_a}) the final output layer exponentiates complex-valued feature maps composed of two real-valued feature maps, calculates sum of these exponentiated values, and projects the sum on a particular irreducible representation of the symmetry group $G$ that corresponds to the symmetry properties of the ansatz wavefuntion:
\begin{equation}\label{eq:ansatz_GCNN_a}
\psi(\vec{\sigma})=\sum_{\hat{g}\in G}\chi_g^*\sum_{n=1}^{N_f}\exp\left(f^{N_l}_{n,g}\right)
\end{equation}
where $N_f$ is the number of complex-valued feature maps (composed of $2\cdot N_f$ real-valued feature maps) in the $N_l$-th layer and $\chi_g$ are characters of the irreducible representation.\cite{Heine,Dresselhaus} The nonlinear activation function before the output is taken to be identity function. 

Within group theory representations are eigenfunctions of the symmetry group operations. Irreducible representations are eigenfunctions of all group operations whilst reducible representations are eigenfunctions only of a subset of these operations. Each irreducible representation has different set of eigenvalues. These eigenvalues are characters $\chi_g$ in the Eq. (\ref{eq:ansatz_GCNN_a}). For non-degenerate symmetry groups where every irreducible representation consists of one eigenfunction, characters are $+1$ or $-1$, whilst for degenerate symmetry groups characters can differ from $\pm 1$.

Described GCNN models are significantly more expressive and yield more accurate results than standard symmetry-averaged models obtained by applying the projection formula: \cite{Heine,Dresselhaus}
\begin{eqnarray}\label{eq:symmetry_average_a}
\left |\psi \right \rangle&=&\frac{d_{\chi}}{|G|}\sum_{\hat{g}\in G} \chi_g^*\hat{g}|\;\tilde{\psi}\;\rangle, \\
\psi\left(\vec{\sigma}\right)&=&\frac{d_{\chi}}{|G|}\sum_{\hat{g}\in G} \chi_g^*\tilde{\psi}\left(\hat{g}^{-1}\vec{\sigma}\right),\nonumber
\end{eqnarray}
where $|G|$ is the order of the group $G$ (the number of its elements) and $d_{\chi}$ is dimension of the irreducible representation with characters $\chi_g$. The projection symmetry averaging approach has been applied to various ansatz wavefunctions\cite{Misawa,Nomura2,Nomura3,Reh,Choo3} and requires evaluation of the model $\tilde{\psi}$ for all symmetry related basis states which can be very computationally expensive. These models are also equivariant with respect to the symmetry group $G$ which can be seen by rewriting an ansatz of the form  (\ref{eq:symmetry_average_a}) as an equivariant model with the output dimension $|G|$:
\begin{eqnarray}\label{eq:equivariant_average_a}
\psi_g\left(\vec{\sigma}\right)&\equiv&\tilde{\psi}\left(\hat{g}^{-1}\vec{\sigma}\right),\\
\psi_g\left(\hat{h}^{-1}\vec{\sigma}\right)&=&\tilde{\psi}\left(\hat{g}^{-1}\hat{h}^{-1}\vec{\sigma}\right)=\psi_{hg}\left(\vec{\sigma}\right).\nonumber
\end{eqnarray}
Within this formulation all $\psi_g\left(\vec{\sigma}\right)$ can be obtained by evaluating the equivariant model for a single input and symmetrization is equivalent to averaging over the model output which is computationally preferable.

For example, for the simpler square lattice with periodic boundary conditions the projection formula (\ref{eq:symmetry_average_a}) becomes familiar Fourier transform that finds crystal momentum eigenstates:
\begin{equation}\label{eq:Fourier_transform_a}
\psi\left(\vec{\sigma}\right)=\frac{1}{L_x\cdot L_y} \sum_{\vec{r}} e^{-i\vec{k}\cdot\vec{r}} \psi_{\vec{r}}\left(\vec{\sigma}\right),
\end{equation}
with $L_x$ and $L_y$ being the lattice dimensions and $\vec{k}=2\pi\cdot\left(n_x/L_x,n_y/L_y\right)$, $n_q \in \left\{0,...,L_q-1\right\}$, $q\in\left\{x,y\right\}$. Irreducible representations of the translation group are therefore one-dimensional and can be labeled by the phases $e^{i\vec{k}\cdot\vec{r}}$.

GCNNs can be mapped to other deep neural network models with symmetry averaging by masking filters, i.e. by setting various off-diagonal feature-to-feature filters (convolution kernels $W$) to zero whilst keeping the input-to-feature filters (embedding kernels $K$) the same. \cite{Roth,Roth2} Diagonal filters are filters with $\hat{h}=\hat{g}$ in (\ref{eq:GCNN_W}) and off-diagonal filters are obtained for $\hat{h}\neq\hat{g}$. For example translationally equivariant CNN model symmetry averaged over a lattice point group symmetries can be obtained by masking off-diagonal filters that correspond to point group rotations and reflections. This model would be more restrictive than a GCNN model equivariant with respect to all space-group symmetries yielding less accurate results.\cite{Cohen,Roth,Roth2} 

It should be noted that the described procedure of constructing the wave-function ans\"atze equivariant with respect to a discrete symmetry group cannot straightforwardly be generalized to continuous symmetry groups like $SU(2)$ spin rotation symmetry. Finding efficient ways to introduce continuous symmetries is still an open research problem with one of the possible solutions being sampling in the basis of irreducible representations instead of spins.\cite{Vieijra,Vieijra2,Misawa}

\section{Training scheme and wave-function optimization}
\label{Appendix_SR}
 The goal of VMC is to find the optimal set of (complex) parameters $\vec{\alpha}=\left\{\alpha_1,...,\alpha_{N_p}\right\}$ in an ansatz wavefunction of the form (\ref{eq:ansatz1_a}) that minimizes a suitably chosen loss function. The standard optimization technique is the gradient descent (GD) method within which the parameters are updated according to the gradients of the chosen loss function with the iterative update rule
\begin{equation}\label{eq:SG_a}
\vec{\alpha}_{t+1}=\vec{\alpha}_t-\eta\vec{\nabla}_{\vec{\alpha}}\mathcal{L},
\end{equation}
where $t$ denotes iteration, $\eta$ is the learning rate (the step size) and $\mathcal{L}$ is a loss function. All numerical methods that attempt to find a global minimum using local information for iterative parameter updates generally exhibit problems related to the local minima in the loss function landscape and slow convergence. Various stochastic methods are therefore introduced to help mentioned numerical algorithms to escape local minima and improve convergence, for example Adam\cite{Kingma} and natural gradient descent (NGD) method.\cite{Amari,Park,Martens,Dong,Park2,Stokes2}

In the Eq. (\ref{eq:SG_a}) -$\vec{\nabla}_{\vec{\alpha}}\mathcal{L}$ represent directions in which the parameters can be updated to obtain the largest decrease of the loss function. The parameters are adjusted by a small step length, i.e. moved for a small distance in the parameter space, in the direction determined by the gradient to optimize the loss function. The standard GD and NGD methods differ by how a small distance is defined. Whilst in the standard GD distance is Euclidean distance that corresponds to the flat metric on the parameter manifold, NGD replaces the flat metric by a generally non-flat metric. NGD method takes into account that changes of all parameters can not be treated equally and that it is not always correct to define distance in terms of how much the parameters are adjusted. Instead a change for each parameter has to be scaled according to the effect of the change on the neural network output distribution. 

To define a new form of distance that accounts for relationships between parameters and reflects changes in the network output distribution NGD method introduces a metric matrix $M$
\begin{equation}\label{eq:NGD_a}
\vec{\alpha}_{t+1}=\vec{\alpha}_t-\eta M^{-1} \vec{\nabla}_{\vec{\alpha}}\mathcal{L},
\end{equation}
In the context of VMC with neural networks NGD method corresponds to the stochastic reconfiguration method (SR),\cite{Carleo,Sorella1,Sorella2,Sorella3,Becca,Stokes2} where the Hilbert-space distance $d$ between two unnormalized wavefunctions $|\phi\rangle$ and $|\phi'\rangle$ is given by the Fubini-Study metrics:
\begin{equation}\label{eq:FS_metrics_a}
\mbox{d}_{FS}=\arccos\sqrt{\frac{\langle\phi'|\phi\rangle\langle\phi|\phi'\rangle}{\langle\phi'|\phi'\rangle\langle \phi|\phi \rangle}}.
\end{equation}
\begin{figure}[t!]
\includegraphics[width=\columnwidth]{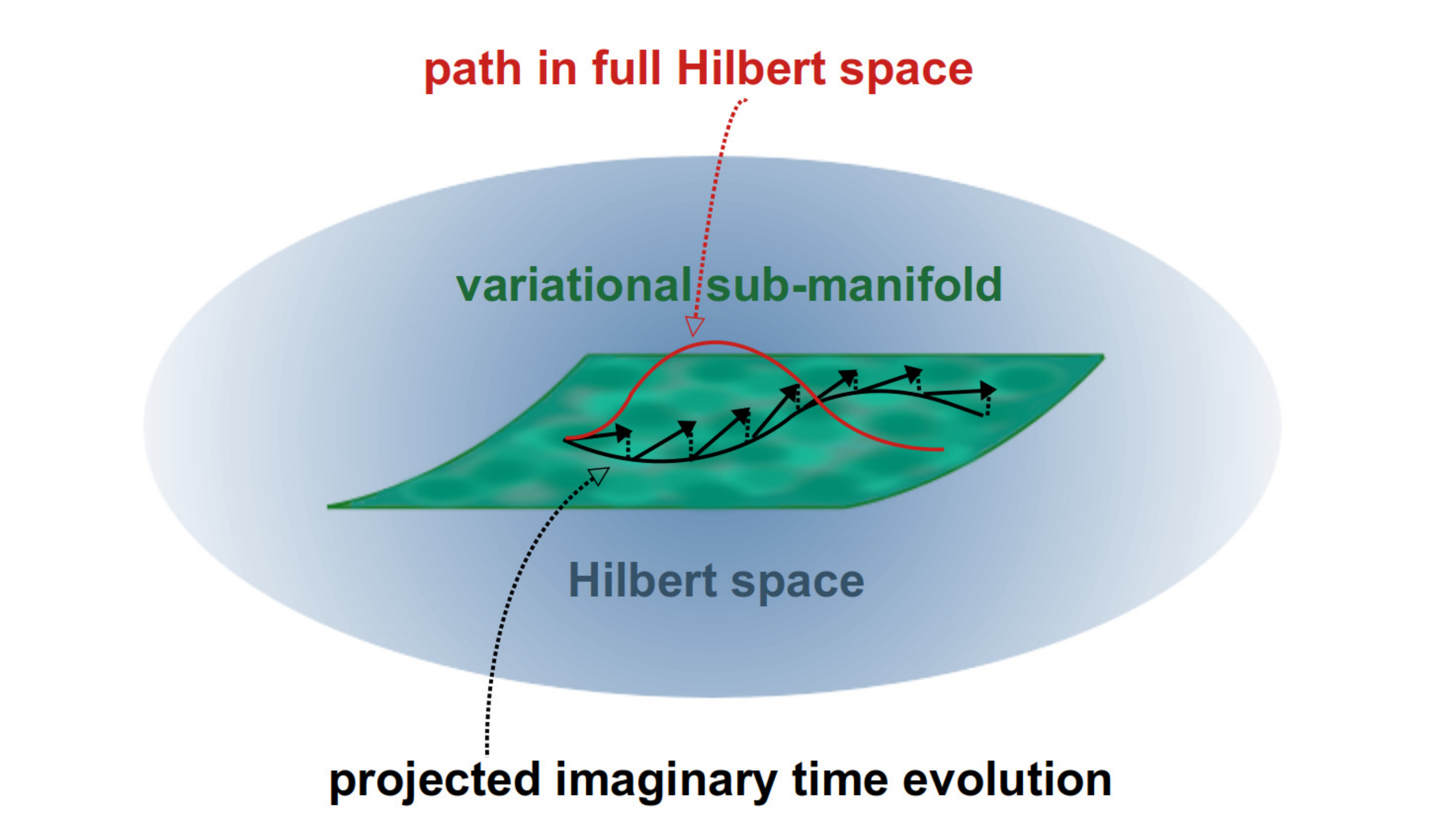}
\caption{\label{Fig:SR} Schematic illustration of the imaginary time evolution in the stochastic reconfiguration method. Starting from a variational wavefunction $|\psi(\vec{\alpha}) \rangle$ with parameters $\vec{\alpha}$ application of the imaginary time evolution operator $e^{-\delta\tau\mathcal{L}}\approx 1 - \Delta \tau\mathcal{L}$ with a small imaginary time step $\Delta\tau$ results in the wavefunction $|\psi'(\vec{\alpha}) \rangle=(1 - \Delta \tau\mathcal{L}) |\psi(\vec{\alpha}) \rangle$ (black arrows). After each imaginary time step $\Delta\tau$ the wavefunction $|\psi'(\vec{\alpha}) \rangle$ is projected to the variational subspace by minimizing the distance $d_{FS}^2$ (\ref{eq:FS_metrics_a}) between $|\psi'(\vec{\alpha}) \rangle$ and $|\psi(\vec{\alpha'}) \rangle$ to obtain new updated set of parameters $\vec{\alpha'}$.
}
\end{figure}
Within SR the parameter updates (\ref{eq:NGD_a}) at each iteration $t$ can be obtained by applying imaginary time evolution operator $e^{-\Delta \tau \mathcal{L}}\approx 1 -\Delta \tau \mathcal{L}$ to the ansatz wavefunction at imaginary time $\tau$:
\begin{equation}\label{eq:imaginary_t_evolution_a}
|\psi'\left(\vec{\alpha}\left(\tau\right)\right)\rangle=\left(1 -\Delta \tau \mathcal{L}\right)|\psi\left(\vec{\alpha}\left(\tau\right)\right)\rangle,
\end{equation}
where $\Delta\tau$ is a small imaginary time step that corresponds to the learning rate $\eta$ in the Eq. (\ref{eq:NGD_a}). After each imaginary time step $\Delta\tau$ the
wavefunction $|\psi'\left(\vec{\alpha}\left(\tau\right)\right)\rangle$ is projected to the variational subspace by minimizing $d_{FS}^2$ with $|\phi\rangle=|\psi\left(\vec{\alpha}\left(\tau\right)\right)\rangle+\Delta\tau\partial_\tau\psi\left(\vec{\alpha}\left(\tau\right)\right)$ and $|\phi'\rangle=|\psi'\left(\vec{\alpha}\left(\tau\right)\right)\rangle$ in the Eq. (\ref{eq:FS_metrics_a}) to obtain a new set of parameters $\vec{\alpha}_{t+1}\equiv\vec{\alpha}\left(\tau+\Delta\tau\right)$:
\begin{equation}\label{eq:SR_update_a}
\vec{\alpha}_{t+1}=\vec{\alpha}_t-\Delta\tau G^{-1} \vec{\nabla}_{\vec{\alpha}}\mathcal{L},
\end{equation}
where the metric matrix $M$ corresponds to the quantum geometric tensor (QGT):
\begin{eqnarray}\label{eq:QGT_a}
&&G_{ij}=\frac{\langle \partial_{\alpha_i}\psi(\vec{\alpha})|\partial_{\alpha_j}\psi(\vec{\alpha})\rangle}{\langle \psi(\vec{\alpha})| \psi(\vec{\alpha})\rangle} \\
&&-\frac{\langle \partial_{\alpha_i}\psi(\vec{\alpha})| \psi(\vec{\alpha})\rangle \langle \psi(\vec{\alpha})|\partial_{\alpha_j} \psi(\vec{\alpha})\rangle}{\langle \psi(\vec{\alpha})|\psi(\vec{\alpha}) \rangle^2},\nonumber
\end{eqnarray}
and $\partial_{\alpha_i}=\partial/\partial \alpha_i$ are partial derivatives with respect to the network parameters $\vec{\alpha}=\left\{\alpha_1,...,\alpha_{N_p}\right\}$. The QGT is also often called the S matrix or quantum Fisher matrix in analogy to the classical Fisher information matrix.\cite{Carleo2,Vicentini2,Stokes2,Amari,Park2} Schematic illustration of the imaginary time evolution in the SR method is shown in FIG. \ref{Fig:SR}.

The loss function in VMC can for example be the variational energy:
\begin{equation}\label{eq:Energy_VMC_a}
\mathcal{L}_E\equiv E= \frac{\langle\psi\left(\vec{\alpha}\right)|H|\psi\left(\vec{\alpha}\right)\rangle}{\langle\psi\left(\vec{\alpha}\right)|\psi\left(\vec{\alpha}\right)\rangle}. 
\end{equation}
where $H$ is the Hamiltonian of the system or free energy at temperature $T$:
\begin{equation}\label{eq:Free_energy_a}
\mathcal{L}_F\equiv F=E-TS,
\end{equation}
where the entropy is defined as\cite{Roth,Roth2,Roth3,Wu,Hibat-Allah2,Hibat-Allah3}
\begin{equation}\label{eq:entropy_a}
S=-\sum_{\vec{\sigma}}\frac{|\psi\left(\vec{\sigma},\vec{\alpha}\right)|^2}{\sum_{\vec{\sigma'}}|\psi\left(\vec{\sigma'},\vec{\alpha}\right)|^2}\log \frac{|\psi\left(\vec{\sigma},\vec{\alpha}\right)|^2}{\sum_{\vec{\sigma'}}|\psi\left(\vec{\sigma'},\vec{\alpha}\right)|^2},
\end{equation}
and $\psi\left(\vec{\sigma},\vec{\alpha}\right)$ is an unnormalized neural-network ansatz wavefunction. 

Minimizing the free energy loss function (\ref{eq:Free_energy_a}) corresponds to minimizing Kullback-Leibler (KL) divergence that measures closeness between two distributions. For a statistical physics model the joint probability for spins $\vec{\sigma}=\left\{\sigma_1,...,\sigma_N\right\}$ follows the Boltzmann distribution 
\begin{equation}\label{eq:Boltzmann_distribution_a}
p\left(\vec{\sigma}\right)=\frac{e^{-\beta E\left(\vec{\sigma}\right)}}{Z},
\end{equation}
where $\beta=1/T$ is the inverse temperature and $Z$ is the partition function and the free energy $\mathcal{F}$ can be obtained as $\mathcal{F}=-\ln Z/\beta$. The KL divergence\cite{Cover}
\begin{equation}\label{eq:KL_divergence_a}
D_{KL}=\sum_{\vec{\sigma}}\psi\left(\vec{\alpha},\vec{\sigma}\right)\ln \left(\frac{\psi\left(\vec{\alpha},\vec{\sigma}\right)}{p\left(\vec{\sigma}\right)}\right)=\beta\left(F-\mathcal{F}\right)
\end{equation}
then measures closeness between the Boltzmann distribution $p(\vec{\sigma})$ and $\psi\left(\vec{\alpha},\vec{\sigma}\right)$ distribution parameterized by variational parameters $\vec{\alpha}$ where 
\begin{equation}\label{eq:Free_energy_2_a}
F=\frac{1}{\beta}\sum_{\vec{\sigma}}\psi\left(\vec{\alpha},\vec{\sigma}\right)\left[\beta E\left(\vec{\sigma}\right)+\ln \psi\left(\vec{\alpha},\vec{\sigma}\right))\right]
\end{equation}
is the variational free energy that gives an upper bound for the free energy $\mathcal{F}$. Since the KL divergence is non-negative, KL minimization is equivalent to the free energy loss (\ref{eq:Free_energy_a}) minimization. 

Within the VMC algorithm the quantum expectation value of any operator $\hat{A}$ 
\begin{equation}\label{eq:Expectation_value_a}
\langle \hat{A}\rangle=\frac{\langle \psi|\hat{A}|\psi\rangle}{\langle \psi|\psi\rangle}
\end{equation}
can be efficiently evaluated by rewriting the expectation value as a Monte Carlo average 
\begin{equation}\label{eq:Expectation_value_MC_a}
\langle \hat{A}\rangle=\sum_{\vec{\sigma}}P\left(\vec{\sigma}\;\right)A^{loc}\left(\vec{\sigma}\;\right)\approx \frac{1}{N_s}\sum_{\vec{\sigma}}A^{loc}\left(\vec{\sigma}\;\right)
\end{equation}
with respect to the quantum probability distribution
\begin{equation}\label{eq:Probability_distribution_a}
P\left(\vec{\sigma}\;\right)=\frac{|\langle \vec{\sigma}\;|\psi\rangle|^2}{\langle \psi|\psi\rangle}
\end{equation}
and estimating the average from $N_s$ samples obtained using Markov chain Monte Carlo (MCMC) sampling methods that generate a sequence of $N_s$ samples that asymptotically follow distribution $P(\vec{\sigma}\;)$, where the local estimator $A^{loc}\left(\vec{\sigma}\;\right)$ is 
\begin{equation}\label{eq:local_estimator_a}
A^{loc}\left(\vec{\sigma}\right)=\frac{\langle \vec{\sigma}\;|\hat{A}|\psi\rangle}{\langle \vec{\sigma}\;|\psi\rangle}=\sum_{\vec{\sigma'}}\frac{\psi(\vec{\sigma'}\;)}{\psi\left(\vec{\sigma}\;\right)}\langle\vec{\sigma}\;|\hat{A}|\vec{\sigma'}\;\rangle.
\end{equation}

Starting from the set of kernels within all layers initialized with Lecun normal initializer that draws samples from a truncated normal distribution centered on zero that gives correct variance of the output,\cite{Klambauer} initial parameter values are updated within the SR algorithm at each iteration according to Eq. (\ref{eq:SR_update_a}) with the free energy loss function described by Eq. (\ref{eq:Free_energy_a}) and the QGT regularization as in Eq. (\ref{eq:QGT_a}). The QGT and the free energy loss function gradient can also be efficiently evaluated as Monte Carlo averages. Whilst the free energy loss in Eq. (\ref{eq:Free_energy_a}) can not be efficiently evaluated from the MCMC samples since our calculations are performed with unnormalized neural-network ansatz wavefunctions and the entropy in Eq. (\ref{eq:entropy_a}) therefore can not be calculated, the loss function gradient within the VMC scheme can be evaluated from
\begin{eqnarray}\label{eq:entropy_gradient_VMC_a}
\vec{\nabla}_{\vec{\alpha}}S&=&-\sum_{\vec{\sigma}}P\left(\vec{\sigma},\vec{\alpha}\;\right)\ln p\left(\vec{\sigma},\vec{\alpha}\;\right)[\;\vec{\nabla}_{\vec{\alpha}}\ln p\left(\vec{\sigma},\vec{\alpha}\;\right) \nonumber\\
&-&\sum_{\vec{\sigma'}}P(\vec{\sigma'},\vec{\alpha}\;)\vec{\nabla}_{\vec{\alpha}}\ln p(\vec{\sigma'},\vec{\alpha}\;)\;]\nonumber\\
&\approx& -\frac{1}{N_s}\sum_{\vec{\sigma}}\ln p\left(\vec{\sigma},\vec{\alpha}\;\right)[\;\vec{\nabla}_{\vec{\alpha}}\ln p\left(\vec{\sigma},\vec{\alpha}\;\right) \nonumber\\
&-&\frac{1}{N_s}\sum_{\vec{\sigma'}}\vec{\nabla}_{\vec{\alpha}}\ln p(\vec{\sigma'},\vec{\alpha}\;)\;],
\end{eqnarray}
where $P\left(\vec{\sigma},\vec{\alpha}\right)$ is the probability distribution (\ref{eq:Probability_distribution_a}), $p\left(\vec{\sigma},\vec{\alpha}\;\right)=|\psi\left(\vec{\sigma},\vec{\alpha}\;\right)|^2$, $\log p=2\mbox{Re}\ln\psi$ and 
\begin{equation}\label{eq:energy_gradient_a}
\vec{\nabla}_{\vec{\alpha}}E=2\mbox{Re}[\vec{f}]
\end{equation}
with $\vec{f}$ being
\begin{eqnarray}\label{eq:derivatives_log_amplitudes_a}
\vec{f}&=&\sum_{\vec{\sigma}} P\left(\vec{\sigma},\vec{\alpha}\;\right)\vec{O}^*\left(\vec{\sigma}\;\right)[\;E^{loc}(\vec{\sigma})\nonumber\\
&-&\sum_{\vec{\sigma'}}P(\vec{\sigma'},\vec{\alpha}\;)E^{loc}(\vec{\sigma'}))\;]\\
&\approx&\frac{1}{N_s}\sum_{\vec{\sigma}}\vec{O}^*\left(\vec{\sigma}\;\right)[\;E^{loc}(\vec{\sigma})-\frac{1}{N_s}\sum_{\vec{\sigma'}}E^{loc}(\vec{\sigma'}))\;]\nonumber
\end{eqnarray}
and $\vec{O}=\vec{\nabla}_{\vec{\alpha}}\ln \psi\left(\sigma,\vec{\alpha}\right)$ the gradient of log-amplitudes. We note that Eq. (\ref{eq:energy_gradient_a}) is valid for non-holomorphic mapping $\vec{\alpha}\rightarrow \psi\left(\vec{\sigma},\vec{\alpha}\right)$ where $\mbox{Re}[\alpha_i]$ and $\mbox{Im}[\alpha_i]$ can be treated as two independent real parameters and Eq. (\ref{eq:energy_gradient_a}) can be applied to each separately.

Additionally correlators of the form  $\langle \hat{A}\hat{B}\rangle$ can be the most efficiently evaluated as covariance of the local estimators for $\hat{A}$ and $\hat{B}$:
\begin{eqnarray}\label{eq:dimer-dimer_cf_VMC_a}
\langle \hat{A}\hat{B}\rangle&=&\frac{\langle \psi|\hat{A}\hat{B}|\psi\rangle}{\langle \psi|\psi\rangle}=\sum_{\vec{\sigma}}P\left(\vec{\sigma}\right)\left[A^{loc}\left(\vec{\sigma}\right)\right]^*B^{loc}\left(\vec{\sigma}\right)\nonumber\\
&\approx&\frac{1}{N_s}\sum_{\vec{\sigma}}\left[A^{loc}\left(\vec{\sigma}\right)\right]^*B^{loc}\left(\vec{\sigma}\right)
\end{eqnarray}
where $P\left(\vec{\sigma}\right)$ is the probability distribution in Eq. (\ref{eq:Probability_distribution_a}).
\end{appendix}


\begin{thebibliography}{99}
\bibitem{Moessner}R. Moessner, and A. P. Ramirez, \textit{Geometrical Frustration}, Physics Today \textbf{59} (2), 24–29 (2006).
\bibitem{Lhuillier}C. Lhuillier, \textit{Frustrated Quantum Magnets}, arXiv:cond-mat/0502464. 
\bibitem{Schmidt}B. Schmidt, P. Thalmeier, \textit{Frustrated two dimensional quantum magnets}, Physics Reports \textbf{703}, 1-59 (2017). 
\bibitem{Balents}L. Balents, \textit{Spin liquids in frustrated magnets}, Nature \textbf{464}, 199 - 208 (2010).
\bibitem{Savary}L. Savary, and L. Balents, \textit{Quantum spin liquids: a review}, Rep. Prog. Phys. \textbf{80}, 016502 (2017).
\bibitem{Broholm}C. Broholm, R. J. Cava, S. A. Kivelson, D. G. Nocera, M. R. Norman, and T. Senthil, \textit{Quantum spin liquids}, Science \textbf{367}, eaay0668 (2020).
\bibitem{Zhou}Y. Zhou, K. Kanoda, and T.- K. Ng, \textit{Quantum spin liquid states}, Rev. Mod. Phys. \textbf{89}, 025003 (2017).
\bibitem{Lee2}P. A. Lee, \textit{From high temperature superconductivity to quantum spin liquid: progress in strong correlation physics}, Rep. Prog. Phys. \textbf{71}, 012501 (2008).
\bibitem{Semeghini}G. Semeghini, H. Levine, A. Keesling, S. Ebadi, T. T. Wang, D. Bluvstein, R. Verresen, H. Pichler, M. Kalinowski, R. Samajdar, A. Omran, S. Sachdev, A. Vishwanath, M. Greiner, V. Vuletic, and M. D. Lukin, \textit{Probing topological spin liquids on a programmable quantum simulator}, Science \textbf{374}, 1242 (2021).
\bibitem{Lauchli}A. M. L\"auchli, J. Sudan, and R. Moessner, \textit{$S=\frac{1}{2}$ kagome Heisenberg antiferromagnet revisited}, Phys. Rev. B \textbf{100}, 155142 (2019).
\bibitem{Wiatek}A. Wietek, and A. M. L\"auchli, \textit{Valence bond solid and possible deconfined quantum criticality in an extended kagome lattice Heisenberg antiferromagnet}, Phys. Rev. B \textbf{102}, 020411(R) (2020).
\bibitem{Changlani2}H. J. Changlani, D. Kochkov, K. Kumar, B. K. Clark, and E. Fradkin, \textit{Macroscopically Degenerate Exactly Solvable Point in the Spin-$\frac{1}{2}$ Kagome Quantum Antiferromagnet}, Phys. Rev. Lett. \textbf{120}, 117202 (2018).
\bibitem{Jiang}H. C. Jiang, Z. Y. Weng, and D. N. Sheng, \textit{Density Matrix Renormalization Group Numerical Study of the Kagome Antiferromagnet}, Phys. Rev. Lett. \textbf{101}, 117203 (2008).
\bibitem{Yan}S. Yan, D. A. Huse, and S. R. White, \textit{Spin-Liquid Ground State of the $S = \frac{1}{2}$ Kagome Heisenberg Antiferromagnet}, Science \textbf{332}, 1173-1176 (2011).
\bibitem{Depenbrock}S. Depenbrock, I. P. McCulloch, and U. Schollw\"ock, \textit{Nature of the Spin-Liquid Ground State of the $S=\frac{1}{2}$ Heisenberg Model on the Kagome Lattice}, Phys. Rev. Lett. \textbf{109}, 067201 (2012).
\bibitem{Jiang2}H.-C. Jiang, Z. Wang, and L. Balents, \textit{Identifying topological order by entanglement entropy}, Nature Physics \textbf{8}, 902-905 (2012).
\bibitem{Nishimoto}S. Nishimoto, N. Shibata, and C. Hotta, \textit{Controlling frustrated liquids and solids with an applied field in a kagome Heisenberg antiferromagnet}, Nature Communications \textbf{4}, 2287 (2013). 
\bibitem{Bauer}B. Bauer, L. Cincio, B. P. Keller, M. Dolfi, G. Vidal, S. Trebst, A. W. W. Ludwig, \textit{Chiral spin liquid and emergent anyons in a Kagome lattice Mott insulator}, Nature Communications \textbf{5}, 5137 (2014).
\bibitem{Gong}S.-S. Gong, W. Zhu, and D. N. Sheng, \textit{Emergent Chiral Spin Liquid: Fractional Quantum Hall Effect in a Kagome Heisenberg Model}, Scientific Reports \textbf{4}, 6317 (2014).
\bibitem{He}Y.- C. He, M. P. Zaletel, M. Oshikawa, and F. Pollmann, \textit{Signatures of Dirac Cones in a DMRG Study of the Kagome Heisenberg Model}, Phys. Rev. X \textbf{7}, 031020 (2017).
\bibitem{Evenbly}G. Evenbly, and G. Vidal, \textit{Frustrated Antiferromagnets with Entanglement Renormalization: Ground State of the Spin-$\frac{1}{2}$ Heisenberg Model on a Kagome Lattice}, Phys. Rev. Lett. \textbf{104}, 187203 (2010).
\bibitem{Xie}Z. Y. Xie, J. Chen, J. F. Yu, X. Kong, B. Normand, T. Xiang, \textit{Tensor Renormalization of Quantum Many-Body Systems Using Projected Entangled Simplex States}, Phys. Rev. X \textbf{4}, 011025 (2014).
\bibitem{Mei}J.-W. Mei, J.-Y. Chen, H. He, and X.-G. Wen, \textit{Gapped spin liquid with topological order for the kagome Heisenberg model}, Phys. Rev. B \textbf{95}, 235107 (2017).
\bibitem{Niu}S. Niu, J. Hasik, J.-Y. Chen, and D. Poilblanc, \textit{Chiral spin liquids on the kagome lattice with projected entangled simplex states}, Phys. Rev. B \textbf{106}, 245119 (2022).
\bibitem{Liao}H. J. Liao, Z. Y. Xie, J. Chen, Z. Y. Liu, H. D. Xie, R. Z. Huang, B. Normand, T. Xiang, \textit{Gapless Spin-Liquid Ground State in the $S=\frac{1}{2}$ Kagome Antiferromagnet}, Phys. Rev. Lett. \textbf{118}, 137202 (2017).
\bibitem{Haghshenas}R. Haghshenas, S.-S. Gong, and D. N. Sheng, \textit{Single-layer tensor network study of the Heisenberg model with chiral interactions on a kagome lattice}, Phys. Rev. B \textbf{99}, 174423 (2019).
\bibitem{Jiang3}S. Jiang, P. Kim, J. H. Han, and Y. Ran, \textit{Competing Spin Liquid Phases in the $S=\frac{1}{2}$
Heisenberg Model on the Kagome Lattice}, SciPost Phys. \textbf{7}, 006 (2019).
\bibitem{Ran}Y. Ran, M. Hermele, P. A. Lee, and X.-G. Wen, \textit{Projected-Wave-Function Study of the Spin-1/2 Heisenberg Model on the Kagom\'e Lattice}, Phys. Rev. Lett. \textbf{98}, 117205 (2007).
\bibitem{Iqbal}Y. Iqbal, F. Becca, and D. Poilblanc, \textit{Valence-bond crystal in the extended kagome spin-$\frac{1}{2}$ quantum Heisenberg antiferromagnet: A variational Monte Carlo approach}, Phys. Rev. B \textbf{83}, 100404(R) (2011).
\bibitem{Clark2}B. K. Clark, J. M. Kinder, E. Neuscamman, G. K.-L. Chan, and M. J. Lawler, \textit{Striped Spin Liquid Crystal Ground State Instability of Kagome Antiferromagnets}, Phys. Rev. Lett. \textbf{111}, 187205 (2013).
\bibitem{Iqbal2}Y. Iqbal, F. Becca, and D. Poilblanc, \textit{Projected wave function study of $\mathbb{Z}_2$ spin liquids on the kagome lattice for the spin-$\frac{1}{2}$ quantum Heisenberg antiferromagnet}, Phys. Rev. B \textbf{84}, 020407(R) (2011).
\bibitem{Tay}T. Tay, and O. I. Motrunich, \textit{Variational study of $J_1-J_2$ Heisenberg model on kagome lattice using projected Schwinger-boson wave functions}, Phys. Rev. B \textbf{84}, 020404(R) (2011).
\bibitem{Hu}W.-J. Hu, W. Zhu, Y. Zhang, S. Gong, F. Becca, and D. N. Sheng, \textit{Variational Monte Carlo study of a chiral spin liquid in the extended Heisenberg model on the kagome lattice}, Phys. Rev. B \textbf{91}, 041124(R) (2015).
\bibitem{Hu2}W.-J. Hu, S.-S. Gong, F. Becca, and D. N. Sheng, \textit{Variational Monte Carlo study of a gapless spin liquid in the spin-$\frac{1}{2}$ XXZ antiferromagnetic model on the kagome lattice}, Phys. Rev. B \textbf{92}, 201105(R) (2015).
\bibitem{Iqbal3}Y. Iqbal, D. Poilblanc, and F. Becca, \textit{Vanishing spin gap in a competing spin-liquid phase in the kagome Heisenberg antiferromagnet}, Phys. Rev. B \textbf{89}, 020407(R) (2014).
\bibitem{Westerhout}T. Westerhout, N. Astrakhantsev, K. S. Tikhonov, M. Katsnelson, and A. A. Bagrov, \textit{Generalization properties of neural network approximations to frustrated magnet ground states}, Nature Communications \textbf{11}, 1593 (2020).
\bibitem{Fu}C. Fu, X. Zhang, H. Zhang, H. Ling, S. Xu, and S. Ji, \textit{Lattice Convolutional Networks for Learning Ground States of Quantum Many-Body Systems}, arXiv:2206.07370. 
\bibitem{Kochkov}D. Kochkov, T. Pfaff, A. Sanchez-Gonzalez, P. Battaglia, B. K. Clark, \textit{Learning ground states of quantum Hamiltonians with graph networks}, arXiv:2110.06390.
\bibitem{Yang}L. Yang, W. Hu, L. Li, \textit{Scalable variational Monte Carlo with graph neural ansatz}, arXiv:2011.12453.
\bibitem{Marston}J. B. Marston, and C. Zeng, \textit{Spin‐Peierls and spin‐liquid phases of Kagom\'e quantum antiferromagnets}, J. Appl. Phys. \textbf{69}, 5962 (1991).
\bibitem{Zeng}C. Zeng, and V. Elser, \textit{Quantum dimer calculations on the spin-1/2 kagom\'e Heisenberg antiferromagnet}, Phys. Rev. B \textbf{51}, 8318 (1995).
\bibitem{Nikolic}P. Nikolic, and T. Senthil, \textit{Physics of low-energy singlet states of the Kagome lattice quantum Heisenberg antiferromagnet}, Phys. Rev. B \textbf{68}, 214415 (2003).
\bibitem{Singh1}R. R. P. Singh, and D. A. Huse, \textit{Ground state of the spin-1/2 kagome-lattice Heisenberg antiferromagnet}, Phys. Rev. B \textbf{76}, 180407(R) (2007).
\bibitem{Singh2}R. R. P. Singh, and D. A. Huse, \textit{Triplet and singlet excitations in the valence bond crystal phase of the kagome lattice Heisenberg model}, Phys. Rev. B \textbf{77}, 144415 (2008).
\bibitem{Poilblanc}D. Poilblanc, M. Mambrini, and D. Schwandt, \textit{Effective quantum dimer model for the kagome Heisenberg antiferromagnet: Nearby quantum critical point and hidden degeneracy}, Phys. Rev. B \textbf{81}, 180402(R) (2010).
\bibitem{Schwandt}D. Schwandt, M. Mambrini, and D. Poilblanc, \textit{Generalized hard-core dimer model approach to low-energy Heisenberg frustrated antiferromagnets: General properties and application to the kagome antiferromagnet}, Phys. Rev. B \textbf{81}, 214413 (2010).
\bibitem{Poilblanc2}D. Poilblanc, and A. Ralko, \textit{Impurity-doped kagome antiferromagnet: A quantum dimer model approach}, Phys. Rev. B \textbf{82}, 174424 (2010).
\bibitem{Hwang}K. Hwang, Y. B. Kim, J. Yu, and K. Park, \textit{Spin cluster operator theory for the kagome lattice antiferromagnet}, Phys. Rev. B \textbf{84}, 205133 (2011).
\bibitem{Poilblanc3}D. Poilblanc, and G. Misguich, \textit{Competing valence bond crystals in the kagome quantum dimer model}, Phys. Rev. B \textbf{84}, 214401 (2011).
\bibitem{Hermele}M. Hermele, Y. Ran, P. A. Lee, and X. G. Wen, \textit{Properties of an algebraic spin liquid on the kagome lattice}, Phys. Rev. B \textbf{77}, 224413 (2008).
\bibitem{Ma}O. Ma and J. B. Marston, \textit{Weak Ferromagnetic Exchange and Anomalous Specific Heat in ZnCu$_3$(OH)$_6$Cl$_2$}, Phys. Rev. Lett. \textbf{101}, 027204 (2008).
\bibitem{Iqbal4}Y. Iqbal, F. Becca, and D. Poilblanc, \textit{Valence-bond crystals in the kagom\'e spin-1/2 Heisenberg antiferromagnet: a symmetry classification and projected wave function study}, New J. Phys. \textbf{14}, 115031 (2012).
\bibitem{Iqbal5}Y. Iqbal, F. Becca, S. Sorella, and D. Poilblanc, \textit{Gapless spin-liquid phase in the kagome spin-$\frac{1}{2}$ Heisenberg antiferromagnet}, Phys. Rev. B \textbf{87}, 060405(R) (2013).
\bibitem{Lee}C.-Y. Lee, B. Normand, and Y.-J. Kao, \textit{Gapless spin liquid in the kagome Heisenberg antiferromagnet with Dzyaloshinskii-Moriya interactions}, Phys. Rev. B \textbf{98}, 224414 (2018).
\bibitem{Pan}G. Pan, and Z. Y. Meng, \textit{Sign Problem in Quantum Monte Carlo Simulation}, The Encyclopedia of Condensed Matter Physics, 2nd edition, Volume 1, Pages 879-893 (2024).
\bibitem{Troyer}M. Troyer and U.-J. Wiese, \textit{Computational Complexity and Fundamental Limitations to Fermionic Quantum Monte Carlo Simulations}, Phys. Rev. Lett. \textbf{94}, 170201 (2005).
\bibitem{White}S. R. White, \textit{Density-matrix algorithms for quantum renormalization groups}, Phys. Rev. B \textbf{48}, 10345 (1993).
\bibitem{White2}S. R. White, \textit{Density matrix formulation for quantum renormalization groups}, Phys. Rev. Lett. \textbf{69}, 2863 (1992).
\bibitem{Schollwock}U. Schollw\"ock, \textit{The density-matrix renormalization group}, Rev. Mod. Phys. \textbf{77}, 259 (2005).
\bibitem{Schollwock2}U. Schollw\"ock, \textit{The density-matrix renormalization group in the age of matrix product states}, Annals of Physics 326, \textbf{96} (2011).
\bibitem{Chan}G. K.-L. Chan, \textit{Low entanglement wavefunctions}, Wiley Interdisciplinary Reviews: Computational Molecular Science, \textbf{2}(6), 907-920 (2012).
\bibitem{Ferrari2}F. Ferrari, A. Parola, and F. Becca, \textit{Gapless spin liquids in disguise}, Phys. Rev. B 103, 195140 (2021).
\bibitem{Hu3}S. Hu, W. Zhu, S. Eggert, and Y.-C. He, \textit{Dirac Spin Liquid on the Spin-1/2 Triangular Heisenberg Antiferromagnet}, Phys. Rev. Lett. \textbf{123}, 207203 (2019).
\bibitem{Zhu}W. Zhu, X. Chen, Y.-C. He, and W. Witczak-Krempa, \textit{Entanglement signatures of emergent Dirac fermions: Kagome spin liquid and quantum criticality}, Science Advances \textbf{4}, 11 (2018).
\bibitem{Carleo2}G. Carleo, K. Choo, D. Hofmann, J. E. T. Smith, T. Westerhout, F. Alet, E. J. Davis, S. Efthymiou, I. Glasser, S.-H. Lin, M. Mauri, G. Mazzola, C. B. Mendl, E. van
Nieuwenburg, O. O'Reilly, H. Th\'eveniaut, G. Torlai, and A. Wietek, \textit{NetKet: A Machine Learning Toolkit for Many-Body Quantum Systems}, SoftwareX \textbf{10}, 100311 (2019).
\bibitem{Vicentini2}F. Vicentini, D. Hofmann, A. Szab\'o, D. Wu, C. Roth, C. Giuliani, G. Pescia, J. Nys, V. Vargas-Calder\'on, N. Astrakhantsev, and G. Carleo, \textit{NetKet 3: Machine Learning Toolbox for Many-Body Quantum Systems}, SciPost Phys. Codebases \textbf{7} (2022); https://www.netket.org/. 
\bibitem{Abadi}M. Abadi, A. Agarwal, P. Barham, E. Brevdo, Z. Chen, C. Citro, G. S. Corrado, A. Davis, J. Dean, M. Devin, S. Ghemawat, I. Goodfellow, A. Harp, G. Irving, M. Isard, Y. Jia, R. Jozefowicz, L. Kaiser, M. Kudlur, J. Levenberg, D. Mane, R. Monga, S. Moore, D. Murray, C. Olah, M. Schuster, J. Shlens, B. Steiner, I. Sutskever, K. Talwar, P. Tucker, V. Vanhoucke, V. Vasudevan, F. Viegas, O. Vinyals, P. Warden, M. Wattenberg, M. Wicke, Y. Yu, and X. Zheng, \textit{TensorFlow: Large-Scale Machine Learning on Heterogeneous Distributed Systems}, 2015 (https://www.tensorflow.org/); arXiv:1603.04467. 
\bibitem{Carleo}G. Carleo, and M. Troyer, \textit{Solving the quantum many-body problem with artificial neural networks}, Science \textbf{355}, 602 (2017).
\bibitem{Cai}Z. Cai, and J. Liu, \textit{Approximating quantum many-body wave functions using artificial neural networks}, Phys. Rev. B \textbf{97}, 035116 (2018).
\bibitem{Liang}X. Liang, W.-Y. Liu, P.-Z. Lin, G.-C. Guo, Y.-S. Zhang, and L. He, \textit{Solving frustrated quantum many-particle models with convolutional neural networks}, Phys. Rev. B \textbf{98}, 104426 (2018).
\bibitem{Choo3}K. Choo, T. Neupert, and G. Carleo, \textit{Two-dimensional frustrated $J_1-J_2$ model studied with neural network quantum states}, Phys. Rev. B \textbf{100}, 125124 (2019).
\bibitem{Ferrari}F. Ferrari, F. Becca, and J. Carrasquilla, \textit{Neural Gutzwiller-projected variational wave functions}, Phys. Rev. B \textbf{100}, 125131 (2019).
\bibitem{Sharir}O. Sharir, Y. Levine, N. Wies, G. Carleo, and A. Shashua, \textit{Deep Autoregressive Models for the Efficient Variational Simulation of Many-Body Quantum Systems}, Phys. Rev. Lett. \textbf{124}, 020503 (2020).
\bibitem{Hibat-Allah}M. Hibat-Allah, M. Ganahl, L. E. Hayward, R. G. Melko, and J. Carrasquilla, \textit{Recurrent neural network wave functions}, Phys. Rev. Research \textbf{2}, 023358 (2020).
\bibitem{Szabo}A. Szab\'o and C. Castelnovo, \textit{Neural network wave functions and the sign problem}, Phys. Rev. Research \textbf{2}, 033075 (2020).
\bibitem{Nomura3}Y. Nomura, and M. Imada, \textit{Dirac-Type Nodal Spin Liquid Revealed by Refined Quantum Many-Body Solver Using Neural-Network Wave Function, Correlation Ratio, and Level Spectroscopy}, Phys. Rev. X \textbf{11}, 031034 (2021).
\bibitem{Astrakhantsev}N. Astrakhantsev, T. Westerhout, A. Tiwari, K. Choo, A. Chen, M. H. Fischer, G. Carleo, and T. Neupert, \textit{Broken-Symmetry Ground States of the Heisenberg Model on the Pyrochlore Lattice}, Phys. Rev. X \textbf{11}, 041021 (2021).
\bibitem{Nomura2}Y. Nomura, \textit{Helping restricted Boltzmann machines with quantum-state representation by restoring symmetry}, J. Phys.: Condens. Matter \textbf{33}, 174003 (2021).
\bibitem{Bukov}M. Bukov, M. Schmitt, and M. Dupont, \textit{Learning the ground state of a non-stoquastic quantum Hamiltonian in a rugged neural network landscape}, SciPost Phys. \textbf{10}, 147 (2021).
\bibitem{Roth} C. Roth, and A. MacDonald, \textit{Group Convolutional Neural Networks Improve Quantum State Accuracy}, arXiv:2104.05085.
\bibitem{Roth2} C. Roth, A. Szab\'o, and A. MacDonald, \textit{High-accuracy variational Monte Carlo for frustrated magnets with deep neural networks}, Phys. Rev. B \textbf{108}, 054410 (2023).
\bibitem{Beck}J. Beck, J. Bodky, J. Motruk, T. M\"uller, R. Thomale, and P. Ghosh, \textit{Phase diagram of the $J-J_d$ Heisenberg model on the maple leaf lattice: Neural networks and density matrix renormalization group}, Phys. Rev. B \textbf{109}, 184422 (2024).
\bibitem{Viteritti}L. L. Viteritti, R. Rende, A. Parola, S. Goldt, and F. Becca, \textit{Transformer Wave Function for two dimensional frustrated magnets: emergence of a Spin-Liquid Phase in the Shastry-Sutherland Model}, arXiv:2311.16889.
\bibitem{Nomura}Y. Nomura, A. S. Darmawan, Y. Yamaji, and M. Imada, \textit{Restricted Boltzmann machine learning for solving strongly correlated quantum systems}, Phys. Rev. B \textbf{96}, 205152 (2017).
\bibitem{Luo}D. Luo, and B. K. Clark, \textit{Backflow Transformations via Neural Networks for Quantum Many-Body Wave Functions}, Phys. Rev. Lett. \textbf{122}, 226401 (2019).
\bibitem{Pfau}D. Pfau, J. S. Spencer, A. G. D. G. Matthews, and W. M. C. Foulkes, \textit{Ab initio solution of the many-electron Schrödinger equation with deep neural networks}, Phys. Rev. Research \textbf{2}, 033429 (2020).
\bibitem{Choo}K. Choo, A. Mezzacapo, and G. Carleo, \textit{Fermionic neural-network states for ab-initio electronic structure}, Nat. Commun. \textbf{11}, 2368 (2020).
\bibitem{Stokes}J. Stokes, J. R. Moreno, E. A. Pnevmatikakis, and G. Carleo, \textit{Phases of two-dimensional spinless lattice fermions with first-quantized deep neural-network quantum states}, Phys. Rev. B \textbf{102}, 205122 (2020).
\bibitem{Cassella}G. Cassella, H. Sutterud, S. Azadi, N. D. Drummond, D. Pfau, J. S. Spencer, and W. M. C. Foulkes, \textit{Discovering Quantum Phase Transitions with Fermionic Neural Networks}, Phys. Rev. Lett. \textbf{130}, 036401 (2023).
\bibitem{Changlani}H. J. Changlani, J. M. Kinder, C. J. Umrigar, and G. Kin-Lic Chan, \textit{Approximating strongly correlated wave functions with correlator product states}, Phys. Rev. B \textbf{80}, 245116 (2009).
\bibitem{Clark}S. R. Clark, \textit{Unifying Neural-network Quantum States and Correlator Product States via Tensor Networks}, J. Phys. A: Math. Theor. \textbf{51} 135301 (2018).
\bibitem{Deng}D.-L. Deng, X. Li, and S. Das Sarma, \textit{Machine learning topological states}, Phys. Rev. B \textbf{96}, 195145 (2017).
\bibitem{Glasser}I. Glasser, N. Pancotti, M. August, I. D. Rodriguez, and J. I. Cirac, \textit{Neural-Network Quantum States, String-Bond States, and Chiral Topological States}, Phys. Rev. X \textbf{8}, 011006 (2018).
\bibitem{Kaubruegger}R. Kaubruegger, L. Pastori, and J. C. Budich, \textit{Chiral topological phases from artificial neural networks}, Phys. Rev. B \textbf{97}, 195136 (2018).
\bibitem{Lu}S. Lu, X. Gao, and L.-M. Duan, \textit{Efficient representation of topologically ordered states with restricted Boltzmann machines}, Phys. Rev. B \textbf{99}, 155136 (2019).
\bibitem{Huang}Y. Huang, and J. E. Moore, \textit{Neural Network Representation of Tensor Network and Chiral States}, Phys. Rev. Lett. \textbf{127}, 170601 (2021).
\bibitem{Vieijra} T. Vieijra, C. Casert, J. Nys, W. De Neve, J. Haegeman, J. Ryckebusch, and F. Verstraete, \textit{Restricted Boltzmann Machines for Quantum States with Non-Abelian or Anyonic Symmetries}, Phys. Rev. Lett. \textbf{124}, 097201 (2020).
\bibitem{Vieijra2}T. Vieijra, and J. Nys, \textit{Many-body quantum states with exact conservation of non-Abelian and lattice symmetries through variational Monte Carlo}, Phys. Rev. B \textbf{104}, 045123 (2021).
\bibitem{Duric}T. \DJ uri\'c, and T. \v{S}eva, \textit{Efficient neural-network based variational Monte Carlo scheme for direct optimization of excited energy states in frustrated quantum systems}, Phys. Rev. B \textbf{102}, 085104 (2020).
\bibitem{Choo2}K. Choo, G. Carleo, N. Regnault, and T. Neupert, \textit{Symmetries and Many-Body Excitations with Neural-Network Quantum States}, Phys. Rev. Lett. \textbf{121}, 167204 (2018).
\bibitem{Schmitt}M. Schmitt, and M. Heyl, \textit{Quantum Many-Body Dynamics in Two Dimensions with Artificial Neural Networks}, Phys. Rev. Lett. \textbf{125}, 100503 (2020).
\bibitem{Gutierrez}I. L. Guti\'errez, and C. B. Mendl, \textit{Real time evolution with neural-network quantum states}, Quantum \textbf{6}, 627 (2022). 
\bibitem{Hofmann}D. Hofmann, G. Fabiani, J. H. Mentink, G. Carleo, and M. A. Sentef, \textit{Role of stochastic noise and generalization error in the time propagation of neural-network quantum states}, SciPost Phys. \textbf{12}, 165 (2022).
\bibitem{Nagy}A. Nagy, and V. Savona, \textit{Variational Quantum Monte Carlo Method with a Neural-Network Ansatz for Open Quantum Systems}, Phys. Rev. Lett. \textbf{122}, 250501 (2019).
\bibitem{Hartmann}M. J. Hartmann, and G. Carleo, \textit{Neural-Network Approach to Dissipative Quantum Many-Body Dynamics}, Phys. Rev. Lett. \textbf{122}, 250502 (2019).
\bibitem{Vicentini}F. Vicentini, A. Biella, N. Regnault, and C. Ciuti, \textit{Variational Neural-Network Ansatz for Steady States in Open Quantum Systems}, Phys. Rev. Lett. \textbf{122}, 250503 (2019).
\bibitem{Yoshioka}N. Yoshioka, and R. Hamazaki, \textit{Constructing neural stationary states for open quantum many-body systems}, Phys. Rev. B \textbf{99}, 214306 (2019).
\bibitem{Luo4}D. Luo, Z. Chen, J. Carrasquilla, and B. K. Clark, \textit{Autoregressive Neural Network for Simulating Open Quantum Systems via a Probabilistic Formulation}, Phys. Rev. Lett. \textbf{128}, 090501 (2022).
\bibitem{Reh2}M. Reh, M. Schmitt, and M. G\"arttner, \textit{Time-Dependent Variational Principle for Open Quantum Systems with Artificial Neural Networks}, Phys. Rev. Lett. \textbf{127}, 230501 (2021).
\bibitem{Irikura}N. Irikura, and Hiroki Saito, \textit{Neural-network quantum states at finite temperature}, Phys. Rev. Research \textbf{2}, 013284 (2020).
\bibitem{Nomura4}Y. Nomura, N. Yoshioka, and F. Nori, \textit{Purifying Deep Boltzmann Machines for Thermal Quantum States}, Phys. Rev. Lett. \textbf{127}, 060601 (2021).
\bibitem{Mizusaki}T. Mizusaki and M. Imada, \textit{Quantum-number projection in the path-integral renormalization group method}, Phys. Rev. B \textbf{69}, 125110 (2004).
\bibitem{Misawa}T. Misawa, S. Morita, K. Yoshimi, M. Kawamura, Y. Motoyama, K. Ido, T. Ohgoe, M. Imada, and T. Kato, \textit{mVMC - Open-source software for many-variable variational Monte Carlo method}, Comp. Phys. Commun. \textbf{235}, 447 (2019).
\bibitem{Seki}K. Seki, T. Shirakawa, and S. Yunoki, \textit{Symmetry-adapted variational quantum eigensolver}, Phys. Rev. A \textbf{101}, 052340 (2020).
\bibitem{Reh}M. Reh, M. Schmitt, and M. G\"arttner, \textit{Optimizing design choices for neural quantum states}, Phys. Rev. B \textbf{107}, 195115 (2023).
\bibitem{Luo2}D. Luo, Z. Chen, K. Hu, Z. Zhao, V. M. Hur, and B. K. Clark, \textit{Gauge-invariant and anyonic-symmetric autoregressive neural network for quantum lattice models}, Phys. Rev. Research \textbf{5}, 013216 (2023).
\bibitem{Luo3}D. Luo, G. Carleo, B. K. Clark, and J. Stokes, \textit{Gauge Equivariant Neural Networks for Quantum Lattice Gauge Theories}, Phys. Rev. Lett. \textbf{127}, 276402 (2021).
\bibitem{Bronstein} M. M. Bronstein, J. Bruna, T. Cohen, and P. Veli\v ckovi\'c, \textit{Geometric Deep Learning: Grids, Groups, Graphs, Geodesics, and Gauges}, arXiv:2104.13478 (2021).
\bibitem{Cohen}T. Cohen and M. Welling, \textit{Group equivariant convolutional networks}, Proceedings of the International Conference on Machine Learning (ICML) (2016); arXiv:1602.07576.
\bibitem{Frostig}R. Frostig, M. J. Johnson, and C. Leary, \emph{Compiling machine learning programs via high-level tracing}, Systems for Machine Learning (2018).
\bibitem{Heek}J. Heek, A. Levskaya, A. Oliver, M. Ritter, B. Rondepierre, A. Steiner, and M. van Zee, \textit{Flax: A neural network library and ecosystem for JAX}, http://github.com/google/
flax (2020).
\bibitem{Hessel}M. Hessel, D. Budden, F. Viola, M. Rosca, E. Sezener, and T. Hennigan, \textit{Optax: composable gradient transformation and optimisation, in jax!}, http://github.com/
deepmind/optax (2020).
\bibitem{Agterberg}D. F. Agterberg, J.C. S\'{e}amus Davis, S. D. Edkins, E. Fradkin, D. J. Van Harlingen, S. A. Kivelson, P. A. Lee, L. Radzihovsky, J. M. Tranquada, and Y. Wang, Annual Review of Condensed Matter Physics, \textit{The Physics of Pair-Density Waves: Cuprate Superconductors and Beyond}, Vol. 11:231-270 (2020).
\bibitem{Lee3}P. A. Lee, \textit{Amperean Pairing and the Pseudogap Phase of Cuprate Superconductors}, Phys. Rev. X \textbf{4}, 031017 (2014).
\bibitem{Lee4}S.- S. Lee, P. A. Lee, and T. Senthil, \textit{Amperean Pairing Instability in the U(1) Spin Liquid State with Fermi Surface and Application to $\kappa$-(BEDT-TTF)$_2$Cu$_2$(CN)$_3$}, Phys. Rev. Lett. \textbf{98}, 067006 (2007).
\bibitem{Xu} X. Y. Xu, K. T. Law, and P. A. Lee, \textit{Pair Density Wave in the Doped $t-J$ Model with Ring Exchange on a Triangular Lattice}, Phys. Rev. Lett. \textbf{122}, 167001 (2019).
\bibitem{Chakraborty}D. Chakraborty, and A. M. Black-Schaffer, \textit{Odd-frequency pair density wave correlations in underdoped cuprates}, New J. Phys. \textbf{23}, 033001 (2021).
\bibitem{LeCun} Y. LeCun, Y. Bengio, and G. Hinton, \textit{Deep learning}, Nature \textbf{521}, 436 (2015).
\bibitem{Goodfellow}I. Goodfellow, Y. Bengio, and A. Courville, \textit{Deep learning} (MIT press, Cambridge, MA, 2016).
\bibitem{Kelleher}John D. Kelleher, \textit{Deep learning} (MIT press, Cambridge, MA, 2019).
\bibitem{Kondor} R. Kondor, and S. Trivedi, \textit{On the generalization of equivariance and convolution in neural networks to the action of compact groups}, Proceedings of the 35th International Conference on Machine Learning (ICML)(2018); arXiv:1802.03690. 
\bibitem{Sorella1}S. Sorella, \textit{Green Function Monte Carlo with Stochastic Reconfiguration}, Phys. Rev. Lett. \textbf{80}, 4558 (1998).
\bibitem{Sorella2}S. Sorella, \textit{Generalized Lanczos algorithm for variational quantum Monte Carlo}, Phys. Rev. B \textbf{64}, 024512 (2001).
\bibitem{Sorella3}S. Sorella, M. Casula, and D. Rocca, \textit{Weak binding between two aromatic rings: Feeling the van der Waals attraction by quantum Monte Carlo methods}, J. Chem. Phys. \textbf{127}, 014105 (2007).
\bibitem{Becca}F. Becca and S. Sorella, \textit{Quantum Monte Carlo Approaches for Correlated Systems} (Cambridge University Press, Cambridge, 2017).
\bibitem{Roth3}C. Roth, \textit{Iterative Retraining of Quantum Spin Models Using Recurrent Neural Networks}, arXiv:2003.06228.
\bibitem{Wu}D. Wu, L. Wang, and P. Zhang, \textit{Solving Statistical Mechanics Using Variational Autoregressive Networks}, Phys. Rev. Lett. \textbf{122}, 080602 (2019).
\bibitem{Hibat-Allah2}M. Hibat-Allah, E. M. Inack, R. Wiersema, R. G. Melko, and J. Carrasquilla, \textit{Variational neural annealing}, Nature Machine Intelligence \textbf{3}, 952–961 (2021).
\bibitem{Hibat-Allah3}M. Hibat-Allah, R. G. Melko, and J. Carrasquilla, \textit{Supplementing Recurrent Neural Network Wave Functions with Symmetry and Annealing to Improve Accuracy}, Machine Learning and the Physical Sciences Workshop (NeurIPS 2021); arXiv:2207.14314.
\bibitem{Kingma}D. P. Kingma and J. Ba, Adam: \textit{A Method for Stochastic Optimization}, 3rd International Conference for Learning Representations (San Diego, 2015); arXiv:1412.6980.
\bibitem{Bamler}R. Bamler, and S. Mandt, \textit{Improving Optimization for Models With Continuous Symmetry Breaking}, Proceedings of the 35th International Conference on Machine Learning (ICML 2018), in PMLR 80:423-432; arXiv:1803.03234.
\bibitem{Aroyo}M. I. Aroyo, \textit{International Tables for Crystallography, Volume A: Space-group Symmetry} (Wiley, New York, 2016).
\bibitem{Gioia}L. Gioia, and C. Wang, \textit{Nonzero Momentum Requires Long-Range Entanglement}, Phys. Rev. X \textbf{12}, 031007 (2022). 
\bibitem{Barkeshli} M. Barkeshli, H. Yao, and S. A. Kivelson, \textit{Gapless spin liquids: Stability and possible experimental relevance}, Phys. Rev. B \textbf{87}, 140402(R) (2013).
\bibitem{Wen}X. - G. Wen, \textit{Quantum orders and symmetric spin liquids}, Phys. Rev. B \textbf{65}, 165113 (2002).
\bibitem{Lu1}Y.- M. Lu, Y. Ran, and P. A. Lee, \textit{$\mathbb{Z}_2$ spin liquids in the $S=\frac{1}{2}$ Heisenberg model on the kagome lattice: A projective symmetry-group study of Schwinger fermion mean-field states}, Phys. Rev. B \textbf{83}, 224413 (2011).
\bibitem{Lu2}Y.- M. Lu, \textit{Symmetry-protected gapless $\mathbb{Z}_2$ spin liquids}, Phys. Rev. B \textbf{97}, 094422 (2018).
\bibitem{Kiese}D. Kiese, F. Ferrari, N. Astrakhantsev, N. Niggemann, P. Ghosh, T. M\"uller, R. Thomale, T. Neupert, J. Reuther, M. J. P. Gingras, S. Trebst, and Y. Iqbal, \textit{Pinch-points to half-moons and up in the stars: The kagome skymap}, Phys. Rev. Research \textbf{5}, L012025 (2023).
\bibitem{Hastings}M. B. Hastings, \textit{Dirac structure, RVB, and Goldstone modes in the kagom\'e antiferromagnet}, Phys. Rev. B \textbf{63}, 014413 (2000).
\bibitem{Iqbal6}Y. Iqbal, D. Poilblanc, and F. Becca, \textit{Spin-$\frac{1}{2}$ Heisenberg $J_1-J_2$ antiferromagnet on the kagome lattice}, Phys. Rev. B \textbf{91}, 020402(R) (2015).
\bibitem{Iqbal7}Y. Iqbal, D. Poilblanc, R. Thomale, and F. Becca, \textit{Persistence of the gapless spin liquid in the breathing kagome Heisenberg antiferromagnet}, Phys. Rev. B \textbf{97}, 115127 (2018).
\bibitem{Fulde}P. Fulde and R. A. Ferrell, \textit{Superconductivity in a Strong Spin-Exchange Field}, Phys. Rev. \textbf{135}, A550 (1964).
\bibitem{Larkin}A. I. Larkin, and Yu. N. Ovchinnikov, \textit{Nonuniform state of superconductors}, Sov. Phys. JETP 20, 762 (1965).
\bibitem{Kitaev}A. Kitaev, \textit{Anyons in an exactly solved model and beyond}, Ann. Phys. \textbf{321}, 2 (2006).
\bibitem{Berg2}E. Berg, C-C. Chen, and S. A. Kivelson, \textit{Stability of Nodal Quasiparticles in Superconductors with Coexisting Orders}, Phys. Rev. Lett. \textbf{100}, 027003 (2008).
\bibitem{Loder}F. Loder, A. P. Kampf, and T. Kopp, \textit{Superconducting state with a finite-momentum pairing mechanism in zero external magnetic field}, Phys. Rev. B \textbf{81}, 020511(R) (2010).
\bibitem{Fradkin}E. Fradkin, S. A. Kivelson, and J. M. Tranquada, \textit{Colloquium: Theory of intertwined orders in high temperature superconductors}, Rev. Mod. Phys. \textbf{87}, 457 (2015).
\bibitem{Berg}E. Berg, E. Fradkin, S. A. Kivelson, and J. Tranquada, \textit{Striped superconductors: how spin, charge and superconducting orders intertwine in the cuprates}, New J. Phys. \textbf{11}, 115004 (2009).
\bibitem{Agterberg2}D. F. Agterbergand, and H. Tsunetsugu, \textit{Dislocations and vortices in pair-density-wave superconductors}, Nature Physics \textbf{4}, 639–642 (2008).
\bibitem{Berg3}E. Berg, E. Fradkin, and S. A. Kivelson, \textit{Charge 4e superconductivity from pair density wave order in certain high temperature superconductors}, Nature Phys. \textbf{5}, 830–833 (2009).
\bibitem{Agterberg3}D. F. Agterberg, and J. Garaud, \textit{Checkerboard order in vortex cores from pair-density-wave superconductivity}, Phys. Rev. B \textbf{91}, 104512 (2015).
\bibitem{Wang2}Y. Wang, S. D. Edkins, M. H. Hamidian, J. C. S\'eamus Davis, E. Fradkin, and S. A. Kivelson, \textit{Pair density waves in superconducting vortex halos}, Phys. Rev. B \textbf{97}, 174510 (2018).
\bibitem{Dai}Z. Dai, Y.-H. Zhang, T. Senthil, and P. A. Lee, \textit{Pair-density waves, charge-density waves, and vortices in high-$T_c$ cuprates}, Phys. Rev. B \textbf{97}, 174511 (2018).
\bibitem{Berg4}E. Berg, E. Fradkin, and S. A. Kivelson, \textit{Theory of the striped superconductor}, Phys. Rev. B \textbf{79}, 064515 (2009).
\bibitem{Radzihovsky}L. Radzihovsky, and A. Vishwanath, \textit{Quantum Liquid Crystals in an Imbalanced Fermi Gas: Fluctuations and Fractional Vortices in Larkin-Ovchinnikov States}, Phys. Rev. Lett. \textbf{103}, 010404 (2009).
\bibitem{Ko}W.-H. Ko, P. A. Lee, and X.-G. Wen, \textit{Doped kagome system as exotic superconductor}, Phys. Rev. B \textbf{79}, 214502 (2009).
\bibitem{Klambauer}G. Klambauer, T. Unterthiner, A. Mayr,and S. Hochreiter, \textit{Self-Normalizing Neural Networks}, Advances in Neural Information Processing Systems 30 (NIPS 2017); 	arXiv:1706.02515.
\bibitem{Heine}V. Heine, \textit{Group Theory in Quantum Mechanics}, (Dover, 2007).
\bibitem{Dresselhaus}M.S. Dresselhaus, G. Dresselhaus, and A. Jorio, \textit{Group Theory - Application to the Physics of Condensed Matter} (Springer-Verlag Berlin Heidelberg, 2008). 
\bibitem{Amari}S.-i. Amari, \textit{Natural gradient works efficiently in learning}, Neural Computation \textbf{10}, 251 (1998).
\bibitem{Park}H. Park, and K. Lee, \textit{Adaptive Natural Gradient Method for Learning of Stochastic Neural Networks in Mini-Batch Mode}, Appl. Sci. \textbf{9}(21), 4568 (2019).
\bibitem{Martens}J. Martens, \textit{New Insights and Perspectives on the Natural Gradient Method}, J. Mach. Learn. Res. \textbf{21}, 146:1-146:76 (2020).
\bibitem{Dong}S. Dong, F. Le, M. Zhang, S.-J. Tao, C. Wang, Y.-J. Han, and G.-P. Guo, \textit{Generalization to the Natural Gradient Descent}, arXiv:2210.02764.
\bibitem{Park2}C.-Y. Park, and M. J. Kastoryano, \textit{Geometry of learning neural quantum states}, Phys. Rev. Research \textbf{2}, 023232 (2020).
\bibitem{Stokes2}J. Stokes, J. Izaac, N. Killoran, and G. Carleo, \textit{Quantum Natural Gradient}, Quantum \textbf{4}, 269 (2020).
\bibitem{Cover}T. M. Cover, and J. A. Thomas, \textit{Elements of information theory} (Wiley, New York, 1991).
\end{thebibliography}
\end{document}